\documentclass[sigplan,10pt]{acmart}
\AtBeginDocument{%
  \providecommand\BibTeX{{%
    \normalfont B\kern-0.5em{\scshape i\kern-0.25em b}\kern-0.8em\TeX}}}

\setcopyright{acmcopyright}
\copyrightyear{2018}
\acmYear{2018}
\acmDOI{XXXXXXX.XXXXXXX}

\acmConference[MPLR]{Make sure to enter the correct
  conference title from your rights confirmation emai}{Oct 22--27,
  2023}{Cascais, Portugal}
\acmBooktitle{Woodstock '18: ACM Symposium on Neural Gaze Detection,
 June 03--05, 2018, Woodstock, NY} 
\acmPrice{15.00}
\acmISBN{978-1-4503-XXXX-X/18/06}

\usepackage{hyperref}

\usepackage{graphicx}
\usepackage{textcomp}
\usepackage{booktabs}
\usepackage{tabularx}
\usepackage{multirow}
\usepackage{multicol}
\usepackage{url}
\usepackage{color,colortbl}
\usepackage{xcolor}

\usepackage{algpseudocode,algorithm,algorithmicx}
\usepackage{listings}
\usepackage{soul} % added by cst
\usepackage{multirow}

\usepackage{caption}
\usepackage{subcaption}

\usepackage[htt]{hyphenat}

\newcommand\NUMBER[1]{#1}

\begin{document}

\title{Comparing Rapid Type Analysis with Points-To Analysis in GraalVM Native
Image}

\author{David Koz\'{a}k}
\email{ikozak@fit.vut.cz}
\affiliation{%
  \institution{Brno University of Technology}
  \city{Brno}
  \country{Czech Republic}
}

\author{Vojin Jovanovic}
\email{vojin.jovanovic@oracle.com}
\affiliation{%
  \institution{Oracle Labs}
  \city{Zurich}
  \country{Switzerland}
}

\author{Codru\c{t} Stancu}
\email{codrut.stancu@oracle.com}
\affiliation{%
  \institution{Oracle Labs}
  \city{Zurich}
  \country{Switzerland}
}

\author{Tom\'{a}\v{s} Vojnar}
\email{vojnar@fit.vut.cz}
\affiliation{%
  \institution{Brno University of Technology}
  \city{Brno}
  \country{Czech Republic}
}

\author{Christian Wimmer}
\email{christian.wimmer@oracle.com}
\affiliation{%
  \institution{Oracle Labs}
  \country{USA}
}

\renewcommand{\shortauthors}{Kozak, Jovanovic, Stancu, Vojnar, and Wimmer.}

\begin{abstract} Whole-program analysis is an essential technique that enables
advanced compiler optimizations. An important example of such a method is
points-to analysis used by ahead-of-time (AOT) compilers to discover program
elements (classes, methods, fields) used on at least one program
path. GraalVM Native
Image uses a points-to analysis to optimize Java applications, which is a
time-consuming step of the build. We explore how much the analysis time can be
improved by replacing the points-to analysis with a rapid type analysis (RTA),
which computes reachable elements faster by allowing more imprecision. We
propose several extensions of previous approaches to RTA: making it parallel,
incremental, and supporting heap snapshotting. We present an extensive
experimental evaluation of the effects of using RTA instead of points-to
analysis, in which RTA allowed us to reduce the analysis time for Spring Petclinic---a~popular demo application of the Spring
framework---by 
\NUMBER{64}\,\% and the overall build time by \NUMBER{35}\,\% at the cost
of increasing the image size due to the imprecision
by~\NUMBER{15}\,\%.\end{abstract}




\received{20 February 2007}
\received[revised]{12 March 2009}
\received[accepted]{5 June 2009}

\maketitle

%==============================================================================
\section{Introduction}
%==============================================================================

Whole-program analysis is an essential technique enabling advanced compiler
optimizations.
An important example of such a technique is \emph{points-to analysis} (PTA)
\cite{pointerAnalysisAreWeThere,dimensionsOfPrecisionInReference,pointerAnalysis,aliasAnalysis}
used to discover program elements (classes, methods, fields) that are used in at
least one run of the program and hence need to be compiled.
We call such elements \textit{reachable}.

\emph{GraalVM Native Image} \cite{native_image_paper} combines PTA,
application initialization at build time, heap snapshotting, and ahead-of-time
(AOT) compilation to optimize Java applications.
This combination of features reduces the application startup time and memory
footprint.
Without using PTA, everything on the Java class path would have to be compiled.
That would lead to long build times and unnecessary large binaries
\cite{native_image_paper}.
The results of the PTA are thus essential, but the overhead of computing
points-to sets for each variable is significant.
It can take minutes to analyze big applications. 

% Moreover, the results of the analysis are used not only
% to reduce the amount of compilation and the size of the resulting binary but
% also to enable more optimizations such as devirtualization, polymorphic inline
% caching, and method inlining \cite{native_image_paper}.

Long build times are inconvenient for developers because they are used to
compiling their applications often, and any delay can significantly hurt their
productivity.
To give a concrete example from our later presented experiments, we can mention
a larger web application (the Spring Petclinic) where the analysis takes
\NUMBER{159} seconds.
We explore how much time can be saved by using a \emph{rapid type analysis}
(RTA) \cite{fast_analysis_cpp, practical_method_resolution_java} instead of PTA.
Intuitively, the basic idea of RTA is to discover which types, i.e., classes
from a given class hierarchy, can be used in methods so-far known to be
reachable from given root methods, which can in turn enlarge the set of
reachable methods by considering that any so-far instantiated type can appear in
a variable of a certain super-type, leading to an iterative fixed point
computation. 

We show how the above idea can be applied in the context of GraalVM Native Image
where one has to deal with issues such as heap snapshotting. 
We build on similar principles as B. Tizer in \cite{virgil}.
On top of that, we also develop a \emph{parallel} and \emph{incremental} version
of the analysis.
The incrementality is achieved by using \emph{method summaries} that sum up the
effect of each analyzed method. These summaries can be serialized and reused
between multiple builds. 

RTA can provide results quicker at the cost of reduced precision.
The lower precision can yield bigger binaries, which is not so problematic
during the development phase.
On the other hand, the need to compile more classes and methods goes against the
savings due to the cheaper analysis.
We try to answer the research question whether such a loss of precision is
justified. 
We perform an extensive experimental comparison of our version of RTA and the
PTA currently implemented in GraalVM Native Image to see whether RTA can provide some advantage
over the PTA, and if so, how much.

For our experiments, we use the standard Java benchmark suites Renaissance
\cite{renaissance} and Dacapo \cite{dacapo} along with example applications for
the Java microservice frameworks Spring \cite{spring}, Micronaut
\cite{micronaut}, and Quarkus \cite{quarkus}.
The experimental evaluation shows, for example, that RTA can reduce the analysis
time of Spring Petclinic---a popular demo application of the Spring
framework---by~\NUMBER{64}\,\% and the overall build time by~\NUMBER{35}\,\% at
the cost of increasing the image size by~\NUMBER{15}\,\%. 
On average, RTA reduced the analysis time by~\NUMBER{40}\,\% and the overall build time by~\NUMBER{15}\,\% 
at the cost of increasing the image size by~\NUMBER{15}\,\%\footnote{Note that
the averages were computed using all our benchmarks including those presented in
the appendix only.}.

We also experiment with the scalability of both RTA and PTA with respect to the
number of available processor cores.
The results show that, for a reduced number of threads, such as \NUMBER{1} or
\NUMBER{4}, the savings in analysis time can be even greater, making RTA a good
choice for constrained environments such as GitHub Actions \cite{githubActions}
or similar CI pipelines.

Our implementation, which is based on the Native Image component of GraalVM~\cite{GraalVM}, is written in Java and Java is used for all examples in this paper. However, our approach is not limited to Java or languages that compile to Java bytecode. It can be applied to all managed languages that are amenable to points-to analysis, such as C\# or other languages of the .NET framework.

In summary, this paper contributes the following:
% \vspace{-5mm}
\begin{itemize}

  \item We introduce a new variant of rapid type analysis for the context of
  GraalVM Native Image. It supports class initialization at build time and heap
  snapshotting.
  
  \item We extend the proposed algorithm to be parallel and incremental.  The
  incrementality is achieved by using \emph{method summaries} that sum up the
  effect of each analyzed method. These summaries can be serialized and reused
  between multiple builds. 

  \item We provide a detailed comparison of the new variant of RTA with a
  points-to analysis for ahead-of-time compilation of Java. We discuss the
  effects on analysis time, build time, reachable elements, and binary size. We
  also evaluate the scalability of both analysis methods. The results show that
  for bigger applications the analysis time can be reduced by up to~\NUMBER{64}\,\%. 

\end{itemize}

% TV: commented out to save space.
%
% The rest of the paper is organized as follows. Sect.~\ref{ni_overview} gives a
% deeper overview of how NI works, focusing specifically on the aspects important
% for the rest of the paper. Sect.~\ref{points_to} describes the usage of PTA in
% NI. Sect.~\ref{analysis_design} describes the extensions and modifications of
% RTA that we propose. Sect.~\ref{evaluation} contains an experimental evaluation.
% Sect.~\ref{state_of_the_art} discusses related works, and, finally,
% Sect.~\ref{conclusion} concludes the paper.

%==============================================================================
\section{Overview of GraalVM Native Image} \label{ni_overview}
%==============================================================================

GraalVM Native Image \cite{native_image_paper} produces standalone binaries for
Java applications that contain the application along with all dependencies and
necessary runtime components such as the garbage collector and threading
support.  It relies on a closed-world assumption, i.e., all code is available to
analyze at image build time.  Dynamic features such as reflection and dynamic
class loading are supported by explicitly registering the program elements that
would otherwise be opaque to the analysis.  The image build process consists of
several successive phases and subphases.

First, the \emph{points-to analysis} is started to detect reachable program
elements.  It starts with a set of \textit{root methods}, which includes the
application entry point specified by the user as well as the entry points of
runtime components. The execution of the analysis is interconnected with
application initialization and heap snapshotting.

\emph{Application initialization} at build time allows developers to initialize
parts of their application when the image is being built instead of performing
the initialization at every application startup. During the initialization,
static fields of initialized classes are assigned to either manually written or
default values.
\emph{Heap snapshotting} traverses all the objects reachable from static fields 
of the initialized classes and constructs an object graph, i.e., a directed 
graph of instances whose edges are references to other 
objects reachable via instance fields or array slots. The object graph constitutes the image
heap and is stored as a part of the binary file. When the application is
started, the image heap is mapped directly into memory
\cite{native_image_paper}. This process and its interaction with static analysis
are discussed in more details in Section~\ref{heap_snapshotting}.

After the analysis finishes, the ahead-of-time compilation is started. 
We use the Graal Compiler
for compilation. Methods are represented using the Graal Intermediate
Representation (IR), which is graph-based and models both the control-flow and
the data-flow dependencies between nodes \cite{graal_ir}. At this point, the IR graphs 
are optimized using the facts proven by the analysis. 
Finally, the image heap and compiled code are written into the image file.

%------------------------------------------------------------------------------
\subsection{Points-to Analysis in GraalVM Native Image} \label{points_to}
%------------------------------------------------------------------------------

This section presents the points-to analysis used in GraalVM Native Image, which
was introduced in \cite{native_image_paper}. The analysis is
%
% by default
%
context-insensitive, path-sensitive, flow-insensitive for fields but
flow-sensitive for local variables. It starts with a set of root methods and
iteratively processes all transitively reachable methods until a fixed-point is
reached.

During the analysis, objects are represented by their types only, not by their
allocation sites as is common in other pointer analyses \cite{pointerAnalysis}.
Using the type abstraction is a sufficiently powerful approximation which yields 
good results in practice when the goal is to compute reachable program elements, 
while keeping the analysis time reasonably low.
This type information is enough to enable compiler optimizations 
such as virtual method de-virtualization.

Each reachable method is parsed from bytecode into the Graal IR, which is then
transformed into a \textit{type-flow graph}.
Nodes of type-flow graphs include those representing instructions as well
as nodes representing \emph{formal parameters} and \emph{return values} of
methods.
The nodes are connected via directed \textit{use} edges.
%
%\st{In case a node has multiple incoming use edges, they are sorted so that
%the different roles of the input nodes are preserved.}
%\cst{This is not true for the type-flow graphs, but only for some of the IR nodes, I think it can be removed. For more details see commented out text.
% More context: In the type-flow graph the role and behavior of a node is determined only by its type, 
% e.g., a filter flow models an instanceof bytecode by only allowing some types through, 
% an invoke node does method resolution and parameter linking.
% The inputs don't need to be disambiguated, they simply propagate their type state to the uses.
% There are a few more intricacies of the type-flow graph, but I don't think we want to go that deep here,
% unless there's enough space left at the end.
%}

Each node maintains a \textit{type-state} information about all types that can
reach it. \emph{Allocation nodes}, i.e., nodes representing allocation
instructions, act as sources that produce types, which are then propagated along
the use edges (with the input/output of nodes representing method invocations
handled in a different way as discussed below). Once a type is added to a
type-state, it is never removed. Thus, the size of all type states can only
grow. In any compilation run, the number of program elements (classes, methods,
and fields) is given and finite. As the number of reachable elements only grows
during the analysis and there is a fixed upper-bound, termination is guaranteed.
The worst-case scenario happens when all program elements are reachable by the
analysis.
%
% : even then the analysis terminates once it marks all elements.

Type-flow graphs of methods are connected into a single interprocedural graph
covering the whole application. For that, nodes producing arguments of method
calls are connected with formal-parameters nodes of the target methods, and
return nodes from the target methods are connected back into the invocation
nodes in the callers. The input edges of invocation nodes are thus not used for
regular type propagation but rather for steering the interconnection of sources
of arguments with the formal-parameter nodes (and of the appropriate return node
with the invocation node).

For \texttt{static} methods, this linkage happens when the type-flow of the
caller is created. For virtual methods, the linkage happens dynamically during
the analysis. Every time a new type is added into the type-state of a receiver
of a method call, it is used to resolve, i.e., to identify, the concrete method
to be linked.

\begin{figure}[t]
\includegraphics[width=6.8cm]{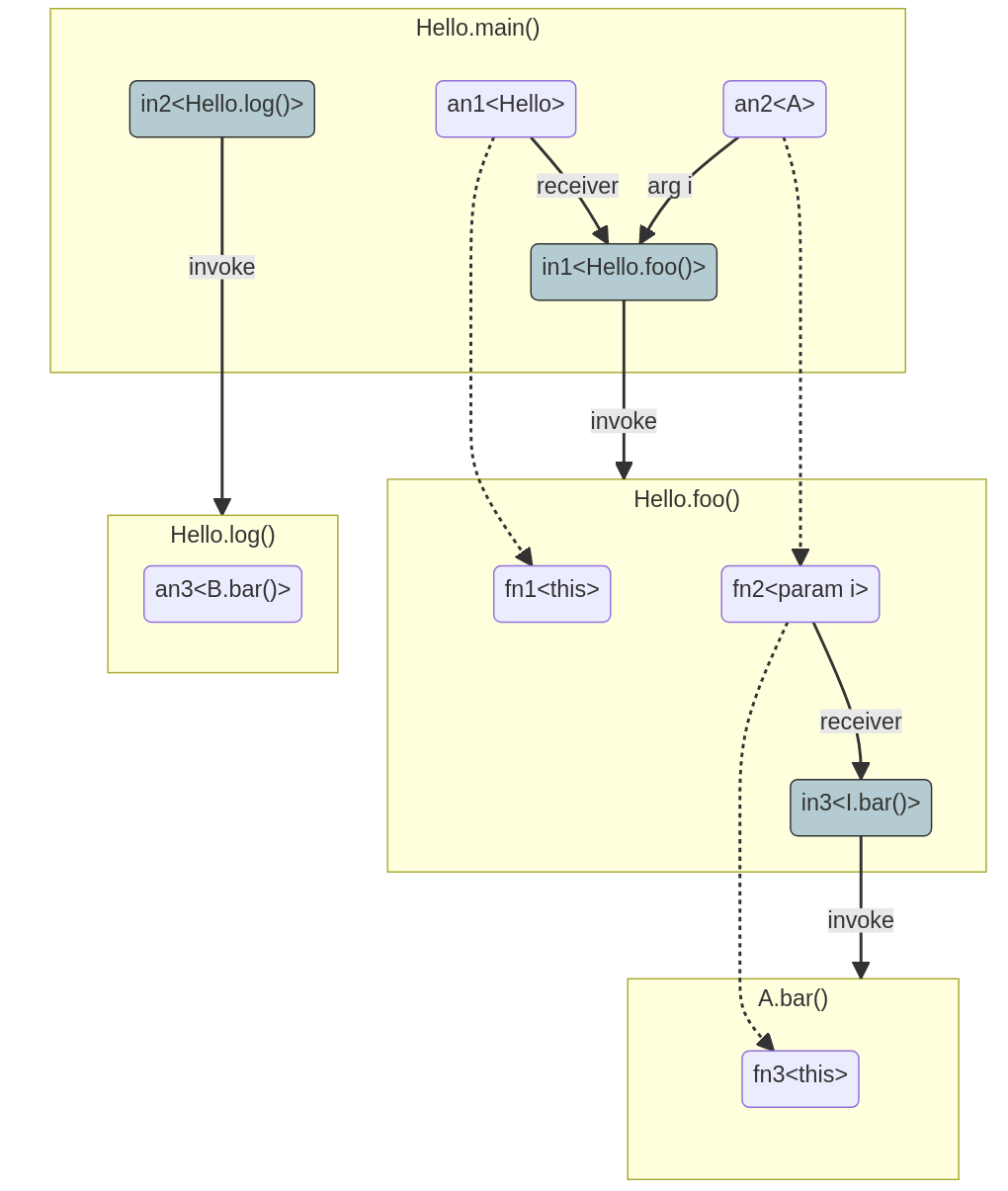}
\centering
\caption{Running example type-flow graph.}
% \vspace{-5mm}
\label{running-example-figure}
\end{figure}

To better understand our PTA, let us now walk through an example. For
brevity, we omit calls to constructors and exception handling.  Consider the
program in Figure~\ref{running-example}.
The analysis starts with the entry point \texttt{Hello.main()}. The method is
parsed, and the type-flow graph in Figure~\ref{running-example-figure} created.
It contains the following nodes:\begin{itemize}

  \item An invocation node $\mathit{in}_1$ for the call of \texttt{foo()}.

  \item An allocation node $\mathit{an}_1$ for \texttt{Hello}, connected to
  $in_1$ as a source of receiver types of the call of \texttt{foo()}.
  
  % TV: I'm guessing there is no node such as this one. I hope I am correct.
  %
  % \item An actual argument node for the parameter \texttt{i} of the invocation
  % of \texttt{Hello.foo}.
  
  % TV: I'm guessing the allocation node is directly connected to the invocation
  % node. I hope this is correct.
  %
  % \item An allocation node for \texttt{A} connected to the actual argument
  % node for \texttt{i}.
  
  \item An allocation node $\mathit{an}_2$ for \texttt{A}, connected to
  $\mathit{in}_1$ as a source of argument types in the call of \texttt{foo()}.
  
  \item An invocation node $\mathit{in}_2$ for the call of \texttt{log()}.    

\end{itemize}

Since $\mathit{an}_1$ is used by $\mathit{in}_1$ as a source of its receiver
types and since the invocation is virtual, as soon as the type \texttt{Hello}
appears in $\mathit{an}_1$, the resolution of virtual methods is used, and
\texttt{Hello.foo()} is found as the method to be invoked.
The body of \texttt{Hello.foo()} is parsed and transformed into the
corresponding type-flow graph with the following nodes:\begin{itemize}

  \item A formal-parameter node $\mathit{fn}_1$ used as a source of types of the
  implicit \texttt{this} parameter.

  \item A formal-parameter node $\mathit{fn}_2$ used as a source of types of the
  formal parameter \texttt{i}.

  \item An invocation node $\mathit{in}_3$ for the call of \texttt{bar()} that
  uses $\mathit{fn}_2$ as a source of its receiver types.

\end{itemize}
Now, $\mathit{an}_1$ and $\mathit{an}_2$ get connected to $\mathit{fn}_1$
and $\mathit{fn}_2$, resp., allowing a flow of types from
\texttt{Hello.main()} into \texttt{Hello.foo()}.
The type \texttt{A} can hence flow from $\mathit{an}_2$ to $\mathit{fn}_2$ and
be used as a receiver type of $\mathit{in}_3$ that is constructed for the call
\texttt{I.bar()}, upon which the resolution selects \texttt{A.bar()} as the call
target.

The call to \texttt{Hello.log()} is static, and so its target can be determined
directly. Its type flow-graph contains an allocation node $\mathit{an}_3$ of
\texttt{B}. Note that while the type \texttt{B} is instantiated, its method
\texttt{B.bar()} is not considered reachable as $\mathit{an}_3$ has no use edge,
and so it can never flow out of the method and get into the invocation node of
\texttt{I.bar()} in \texttt{Hello.foo()}.

\begin{figure}[tbp]
\begin{lstlisting}
public class Hello {
    public static void main() {
        new Hello().foo(new A());
        log();
    }
    static void log() { new B(); }   
    void foo(I i) { i.bar(); }
}
interface I { void bar(); }
class A implements I {...}
class B implements I {...}
\end{lstlisting}
% \vspace{-5mm}
\caption{Running example for analysis.}
% \vspace{-5mm}
\label{running-example}
\end{figure}

% %% Mermaid markdown code for the figure
% graph TB;
%     classDef invoke fill:#b4ccd1,stroke:#333,stroke-width:1px;
%     classDef subgraph_padding fill:none,stroke:none %% title padding hack

%     subgraph main ["Hello.main()"]
%         subgraph subgraph_padding1 [ ]
%             in1("in1&lt;Hello.foo()&gt;"):::invoke
%             an1("an1&lt;Hello&gt;") --receiver--> in1
%             an2("an2&lt;A&gt;") --arg i--> in1
%             in2("in2&lt;Hello.log()&gt;"):::invoke
%         end
%     end
   
%     in1 --invoke--> foo
%     an1 -..-> fn1
%     an2 -..-> fn2
%     subgraph foo ["Hello.foo()"]
%         fn1("fn1&lt;this&gt;") 
%         fn2("fn2&lt;param i&gt;")
%         in3("in3&lt;I.bar()&gt;"):::invoke
%         fn2 --receiver--> in3
%     end

%     in2 --invoke---> log
%     subgraph log["Hello.log()"]
%         an3("an3&lt;B.bar()&gt;")
    
%     end

%     in3 --invoke--> aBar
%     fn2 -..-> fn3
%     subgraph aBar["A.bar()"]
%       subgraph subgraph_padding2 [ ]
%             fn3("fn3&lt;this&gt;") 
%         end
%     end

%     class subgraph_padding1 subgraph_padding
%     class subgraph_padding2 subgraph_padding

The results of the points-to analysis are useful not only to identify reachable
elements but also for many compiler optimizations. For example, they can be used
to remove unnecessary casts, remove dead branches of \texttt{instanceof} checks
that are always true or false, exclude fields that are never accessed, and to
optimize virtual calls with a limited number of receiver types. Knowing the set
of receivers and their types allows one to devirtualize method calls with only
one receiver type, employ polymorphic inline caching
\cite{polymorphic_inline_cache} when there are a few receiver types only, and
perform more method inlining, which can lead to subsequent
optimizations. 

%==============================================================================
\section{RTA with Method Summaries} \label{analysis_design}
%==============================================================================

This section presents our implementation of RTA \cite{fast_analysis_cpp,
practical_method_resolution_java}.
It supports heap snapshotting, is designed to be parallel, and supports method
summaries to make it incremental.
First, we describe the basic idea of the core of the analysis using a~system of high-level
constraints, which neglects some more technical aspects of the actual analysis
to be easier to understand.
Then, we describe a single-threaded, non-incremental version of the proposed
analysis, which, however, already contains some preparation for its subsequent
parallelization.
Afterwards, we propose how to run parts of the analysis in parallel, and,
finally, we discuss incremental analysis.

The basic effect of the analysis---assuming all method calls to be virtual
(i.e., not distinguishing different types of invocations)---can be
summarized using the following constraints inspired by the work of Tip et al. in
\cite{scalablePropagationBasedCallGraphAlgs}.

Let $T$ be the set of all types, $M$ the set of all methods, and $E$ the set of
all expressions in the analyzed application.
We use $StaticType(e)$ to denote the static type of $e \in E$.
Furthermore, $Subtypes(t)$ denotes the set of all subtypes of $t \in T$, and
$StaticLookup(t,m)$ denotes the actual call target for a virtually invoked $m
\in M$ on $t \in T$.
For $m \in M$, let $CallExpr(m)$ denote the set of all call expressions $e.f()$
for $e \in E$ and $f \in M$ that appear in the method $m$, and let $InstExpr(m)$
denote the set of all instantiation expressions $\mathtt{new}\,C()$ for $C \in
T$ that appear in $m$.
The sets $R \subseteq M$ and $I \subseteq T$ representing reachable methods and
instantiated types determined using RTA satisfy the following
constraints:\begin{enumerate}

  \item $\mathtt{main} \in R$.

  \item $\forall m \in R ~ \forall e.f() \in CallExpr(m)\\ \forall t \in
  Subtypes(StaticType(e)).$\\ $t \in I \land StaticLookup(t,f) = m' \implies m'
  \in R$.

  \item $\forall m \in R ~ \forall new\,C() \in InstExpr(m). ~ C \in I$.

\end{enumerate}

Intuitively, $\mathtt{main}$ is always reachable.
The second rule makes sure that all methods that can be virtually called from a
call expression in a reachable method are also reachable.
Finally, the third rule makes sure that any type that can be instantiated in a
reachable method is considered instantiated.

%% TV -- too long:
%
% We will now describe each of the rules in more detail. The Rule 1) specifies
% that the main method is contained within $R$. It can be perceived as the
% starting rule, as the other rules further extend the content of $R$ and $I$
% based on what is already contained within $R$. 
% %
% The Rule 2) describes the handling of virtual invokes. For each method $m \in M$
% that is also contained in $R$ (it is already proven reachable) and each virtual
% invoke $e.f()$ within $m$, the set of all subtypes of the static type of $e$ is
% processed and each of the types within the set that is also included in $I$ (it
% is proven instantiated) is queried for its own implementation of $f$ using
% $StaticLookup(t,f)$. The results of the lookup are included in $R$. 
% %
% The Rule 3) handles instantiating new types. It ensures that for each method $m
% \in M$ that is also contained in $R$ (so it is already proven reachable) and for
% each $new\,C()$ in $m$, the type $C$ is included in $I$ (it is proven
% instantiated). 

The above constraints showed how RTA handles virtual method invocations.
However, there are actually five different types of invokes in Java:
\texttt{invokestatic}, \texttt{invokevirtual}, \texttt{invokeinterface},
\texttt{invokespecial}, and \texttt{invokedynamic}.

% \begin{itemize}
% 
%   \item \texttt{invokestatic},
%
%   \item \texttt{invokevirtual},
%
%   \item \texttt{invokeinterface},
%
%   \item \texttt{invokespecial}, and
%
%   \item \texttt{invokedynamic}.
%
% \end{itemize}

As defined in the JVM specification \cite{JVMSpec}, \texttt{invokevirtual} and
\texttt{invokeinterface} represent virtual method invocations, and we do not
need to distinguish them for our purposes. 
\texttt{Invokestatic} represents static method invocation, i.e., a direct
invocation of a method called on a~Java class, not on an instance.
\texttt{Invokespecial} represents a~direct invocation of an instance method in
cases where it is clear which method should be called.
This instruction is used, for example, when calling constructors, when calling a
method on the superclass of the current class, or when calling a method on an
expression of a type that has no subtypes.
Both \texttt{invokestatic} and \texttt{invokespecial} are \textit{direct}
invokes, i.e. they have a unique call target that can be statically determined.
Therefore, they could be resolved in the same way immediately upon the discovery
of the invoke instruction in the bytecode of any reachable method.
However, as shown in Section~\ref{section:invokeSpecial}, differentiating
between them can actually increase the precision in some cases.
\texttt{Invokedynamic} represents a special invoke whose call target is not yet
fixed but computed on the first execution of the bytecode.
As our analysis is based on the Graal IR, it does not have to handle
\texttt{invokedynamic} explicitly because these invokes are processed by the
Graal compiler before the analysis starts and are optimized either into direct
invokes or into lookup procedures determining the correct call target at
runtime.

%
% \tv{??? What do we do with this kind of call? We should say that even if we
% ignore it for some reason. Otherwise, questions will arise.}
%
% \dk{The analysis does need to handle \texttt{invokedynamic} explicitly because it either already sees a 
% concrete call target if it could be determined by the compiler before the analysis 
% starts, or it sees an invocation of a "bootstrap" method responsible for determining the call target.
% However, I am not sure about including this piece of text in the paper, because it is one of the things described in the 
% other analysis paper, so I would prefer to be brief or maybe somehow completely ignore this invoke since we do not really deal with it.
% I've included it to be complete, but it is one of the things I would still like to discuss.}
%
% \tv{Unfortunately, I do not understand the text you have at the beginning of the
% remark. I do not insist we should be too precise here, but (1) we must explain what
% we do with this kind of call, and (2) it must be clear that we handle it in a
% sound way. For me, it is immediately visible that there is something fishy going
% on here.}

% \vspace*{20mm} % For LaTeX to be happy.

%------------------------------------------------------------------------------
\subsection{Core Algorithm and Data Structures} \label{section:coreAlg}
%------------------------------------------------------------------------------

We now refine the above presented basic idea of RTA such that (1) it takes into
account different kinds of calls that can appear (static calls, virtual calls,
special calls), (2) computes information needed for subsequent compilation
phases (in order not to have to repeat the analysis for this purpose), and (3)
is ready for subsequent parallelization.

% Having defined the effect of the analysis, let us now proceed to the
% description of the core data structures and algorithms used to actually
% compute the results.

During the analysis, the effect of each method is represented using a
\emph{method summary} that consists of sets that contain the following
information: static invoked methods, virtually invoked methods, special invoked
methods, instantiated types, read fields, written fields, and
embedded constants.
%
% \begin{itemize}
%  
%   \item static invoked methods,
%  
%   \item virtually invoked methods,
%  
%   \item special invoked methods,
%  
%   \item accessed types,
%  
%   \item instantiated types,
%  
%   \item read fields,
%  
%   \item written fields, and
%  
%   \item embedded constants.
%  
% \end{itemize}
%
The summary format is designed to be minimal while still containing all the
necessary information for both RTA and the later AOT compilation step.
For example, distinguishing between read and written fields is not needed for
RTA itself, but AOT compilation requires the information since it automatically
removes never accessed fields \cite{native_image_paper}.
The information about which fields are read is also needed to drive the heap
snapshotting (cf. Section~\ref{heap_snapshotting}).
%
% the scanner is notified for each read field

% \fixme{(!!!) TV: I got completely lost when reading the pseudocode of the
% analysis since I did not see any effect of many of the methods. The reason is
% that they are updating the state of the analysis that is nowhere explicitly
% mentioned. I am sorry I did not realize this sooner. However, I think the
% description simply cannot stay as it is now. We need to: (1) explain how the
% state of the analysis looks like, (2) introduce two global variables into the
% code---namely, \texttt{worklist} and \texttt{state}, and (3) make the use of
% \texttt{state} as explicit as possible. I hope that can be done even with the
% parallel extension in mind.}

% \todo{The analysis uses two global variables: the worklist consisting of the
% methods to be analysed and the state of the analysis. The state of the analysis
% is ...}

The internal state of the analysis can be viewed as consisting of
a worklist containing all methods that still need to be analyzed and the
following pieces of information associated with the representation kept by the
compiler for types, methods, and fields:
%
%cwimmer: I did not find this footnote necessary and helpful, so I commented it out.
%
%\footnote{This distributed representation
%of the state allows for concurrent access to the information on the particular
%program elements. Moreover, we note that some pieces of information are in fact
%duplicated (e.g., each type is marked as instantiated or not, at the same time
%keeping with it a set of all its so-far discovered instantiated sub-types),
%which is done on purpose to make the later presented parallel version easier to
%implement.}:
%
%
% The rest of the state is distributed into the representations for types,
% methods and fields. 
%
\begin{itemize}

  \item For each \emph{type} \texttt{t}, the analysis stores the following:
  \begin{itemize} 

    % \item Its supertype and implemented interfaces, which allows traversing
    % the type hierarchy upwards.

    \item An atomic boolean flag set to true if the analysis discovers that $t$
    may be \emph{instantiated} at run time.

    \item A~set of methods declared in \texttt{t} that the analysis has
    so-far found to be \emph{virtually invoked}.

    \item A~set of methods declared in \texttt{t} that the analysis has
    so-far found to be \emph{special invoked}.

    \item A~set of subtypes of \texttt{t} discovered as instantiated.

    % Both of the above sets empty when the analysis starts. 

  \end{itemize}

  \item For each \emph{method} \texttt{m}, the analysis stores the
  following:\begin{itemize}

    \item An atomic boolean flag marking $m$ as \emph{invoked}, indicating that
    the method body is considered reachable at runtime.

    \item An atomic boolean flag marking $m$ as \emph{special invoked}, indicating that
    the method may be a target of an invoke special call.

    \item An atomic boolean flag marking $m$ as \emph{virtually invoked}, which
    indicates that the method may be the target of a virtual method call. Note
    that this does not necessarily mean that $m$ is invoked since the invoked
    method can come from some subtype of the declaring type of $m$.

  \end{itemize}

  \item For each \emph{field} \texttt{f}, the analysis stores the following:
  \begin{itemize}

    \item An atomic boolean flag marking $f$ as \emph{read}.

    \item An atomic boolean flag marking $f$ as \emph{written}.

\end{itemize}

\end{itemize}

The pseudocode of the core of the analysis can be found in
Algorithm~\ref{coreAlg}. It starts with a set of root methods used to initialize
the worklist (line 1). The main loop (lines~2--7) then processes methods in the
worklist until it becomes empty. 

For each method in the worklist, it is first parsed into the Graal IR, the
intermediate representation discussed previously (line 4).
The summary of the method being processed is initialized to consist of empty
sets.
The \texttt{extractSummary} method (line 5) then iterates over the instructions
of the method, and whenever it finds an instruction of the types listed in the
left column of Table~\ref{graal_ir_summary_correspondence}, it adds it to the
collection of the summary given in the right column.

\begin{table}[t]
    \centering
    \caption{Correspondence between bytecode instructions and collections in 
      method summaries.}
    % \vspace{-5mm}
    \begin{tabular}{ll}
    \hline
        \toprule
        Bytecode instruction & Collection in the summary \\ 
        \midrule
        new & Instantiated types \\ 
        anewarray & Instantiated types \\ 
        multianewarray & Instantiated types \\ 
        getfield & Read fields \\ 
        putfield & Written types \\ 
        invoke* & Directly/Virtually called methods \\ 
        \bottomrule
    \end{tabular}
    % \vspace{-5mm}
    \label{graal_ir_summary_correspondence}
\end{table}

When the summary is ready, it is passed into the \texttt{appplySummary} method
(lines~9--20).
This method iterates over all collections within the summary and calls
appropriate \texttt{register} methods. 

Many of the \texttt{register} methods are relatively straightforward.
For example, see the method \texttt{registerAsInvoked} (lines~22--26), which
adds an invoked method to the worklist.
Note that the \texttt{mark} method (line~23) is called before adding the method
being processed into the worklist.
The \texttt{mark} method accepts a boolean flag as a parameter.
If the flag is \texttt{true}, \texttt{mark} returns \texttt{false} (intuitively,
no marking was needed).
If the flag is \texttt{false}, \texttt{mark} atomically changes it to
\texttt{true} and returns \texttt{true} (intuitively, the marking was needed).
Hence, \texttt{mark} returns \texttt{true} only on its first invocation and
\texttt{false} otherwise.
This implementation with the atomic update is used to facilitate the parallel
analysis presented later on.
Registering fields as read or written follows the same pattern, and is omitted
from the algorithm for space reasons.

%------------------------------------------------------------------------------
\subsection{Invoke Virtual Handling} \label{section:invokeVirtual}
%------------------------------------------------------------------------------

The handling of virtual invokes and instantiated types is more interesting (see
Algorithm~\ref{rtaMethodsAlg}).
The two methods presented in the algorithm are interconnected. 

The \texttt{registerAsVirtualInvoked} method (lines~1--11) handles a virtual
method call.
Since the analysis has no points-to information, it uses the declaring class of
the invoked method to traverse all its currently instantiated subtypes.
The information about instantiated subtypes is collected inside the
\texttt{registerAsInstantiated} method (line~16).
For each instantiated subtype, the virtual method is resolved into a concrete
method using \texttt{type.resolveMethod} (line~7), which resolves a virtual call
or interface call for the given concrete caller type according to the Java VM
specification~\cite{JVMSpec}.

The \texttt{registerAsInstantiated} method (lines~13--26) is used when a type is
instantiated.
First, the newly instantiated type is added to the \texttt{instantiatedSubytpes}
set of all supertypes.
Then the supertype hierarchy is traversed again, and for each visited type, the
list of all virtually invoked methods is processed (lines~20--23).
This list is collected as a part of \texttt{registerAsVirtualInvoked} (line~4).
For each virtually invoked method in the list, \texttt{type.resolveMethod} is
used to obtain the concrete method (line~21).

Note that essentially the same method resolution is performed by both
\texttt{registerAsVirtualInvoked} and \texttt{registerAsInstantiated} but from
two different perspectives.
It is not possible to have only one of them.
That would require an ordering in which the \texttt{register} methods that only
mark elements are called before the \texttt{register} methods that do the
resolution.
Such an ordering is not possible because discovering instantiated types and
invoked methods is interconnected.
Discovering new instantiated types makes new methods reachable from invokes
within already analyzed methods and vice versa.

\begin{algorithm}
        \begin{flushleft}
          % \hspace*{4pt}
          \textbf{Input:} The set of root methods \\
          % \hspace*{4pt}
          \textbf{Output:} All reachable types, methods, and fields\\
          % \hspace*{4pt}
        %   \chck{\textbf{Globals:} \texttt{worklist}, \texttt{state}}\\ 
        \end{flushleft}
        \begin{algorithmic}[1]
        \State $worklist \gets rootMethods$
        % \\\todo{One should also initialize \texttt{state}}
        \While{$worklist \ne \emptyset$}
        \State $method \gets removeFirst(worklist)$ 
        \State $irGraph \gets parseMethod(method)$
        \State $summary \gets extractSummary(irGraph)$
        \State $applySummary(summary)$ 
        \EndWhile
        % \State
        \Procedure{applySummary}{$summary$}
            \For{$m \gets summary.directInvokes$}
                \State $registerAsInvoked(m)$
            \EndFor
            \For{$m \gets summary.virtualInvokes$}
                \State $registerAsVirtualInvoked(m)$
            \EndFor
            \For{$t \gets summary.instantiatedTypes$}
                \State $registerAsInstantiated(t)$
            \EndFor
            \State $//\;...\;similar\;loops\;for\;other\;parts\;...$
        \EndProcedure
        % \State
        \Procedure{registerAsInvoked}{$m$}
            \If{$mark(m.isInvoked)$}
                \State $worklist.add(m)$
            \EndIf
        \EndProcedure
        % \State
        \Procedure{mark}{$flag$}
            \State \Return $flag.compareAndSet(false,true)$
        \EndProcedure
        
        \end{algorithmic}
        \caption{Rapid type analysis worklist loop.}
        \label{coreAlg}
\end{algorithm}

\begin{algorithm}
        \begin{algorithmic}[1]
        \Procedure{registerAsVirtualInvoked}{$m$}
            \If{$mark(m.isVirtualInvoked)$}
                \State $t \gets m.declaringType$
                \State $t.virtualInvokedMethods.add(m)$
                % \State
                \For{$subt \in t.instantiatedSubtypes$}
                    \State $resolved \gets subt.resolveMethod(m)$
                    \State $registerAsInvoked(resolved)$ 
                \EndFor
            \EndIf
        \EndProcedure
        % \State
        \Procedure{registerAsInstantiated}{$t$}
            \If{$mark(t.isInstantiated)$}
                \For{$supert \in t.superTypes$} 
                    \State $supert.instantiatedSubtypes.add(t)$ 
                \EndFor
                % \State
                \For{$supert \in t.superTypes$}
                    \For{$m \in supert.virtualInvokedMethods$}
                        \State $resolved \gets t.resolveMethod(m)$
                        \State $registerAsInvoked(resolved)$ 
                    \EndFor
                \EndFor
            \EndIf
        \EndProcedure
        \end{algorithmic}
        \caption{RTA handling of virtual methods.}
        \label{rtaMethodsAlg}
\end{algorithm}

%------------------------------------------------------------------------------
\subsection{Invoke Special Handling} \label{section:invokeSpecial}
%------------------------------------------------------------------------------

Another interesting case that we would like to discuss is
\texttt{invokespecial}.
Differentiating static invokes and special invokes is not necessary for
correctness.
However, it increases precision because there are cases when a special invoke is reachable, but no
object is instantiated on which such method can be called.
This happens, for example, when analyzing the
method \texttt{java.lang.Thread.sleep}.
A simplified version of the method for Java 20 can be found in Figure~\ref{java-lang-thread}.
It contains the handling of virtual threads from the project
Loom \cite{projectLoom} (instances of \texttt{VirtualThread}) although their
usage has to be explicitly enabled by a command line option.
Otherwise, no virtual thread is ever created, and the content of the \texttt{if}
block is dead code. 
However, without the special handling described below, the method
\texttt{VirtualThread.sleep} would be considered invoked even when the support
for virtual threads was disabled.

\begin{figure}[htbp]
% \vspace{-3mm}
\begin{lstlisting}
public class Thread {
    public static void sleep(long millis) {
        ...
        if (currentThread() instanceof VirtualThread vthread){
            vthread.sleep(millis);
            return;
        }
        ...
    }
}
\end{lstlisting}
% \vspace{-5mm}
\caption{Invoke special example.}
% \vspace{-4mm}
\label{java-lang-thread}
\end{figure}

Due to the above, we handle \texttt{invokespecial} separately as shown in Algorithm~\ref{invokeSpecialAlg}.
The method \texttt{registerAsSpecialInvoked}
(lines 1--10) performs two tasks:
First, it adds the called method to the set of invoked special methods on the
declaring type (line 4).
Then, it calls \texttt{registerAsInvoked} but only if any subtype of the
declaring type has been instantiated so far (lines 6--8).
Similarly to the previously described handling of virtual methods, it is also
necessary to handle the case where the method is processed first and the type
instantiated later.
Therefore, extend the method \texttt{registerAsInstantiated} with
another loop that iterates over all invoked special methods of all supertypes of
the newly instantiated type and processes them via  \texttt{registerAsInvoked}
(lines 16--18).
This way, we delay the processing of \texttt{invokespecial} only after a
suitable type upon which they can be called has been instantiated.

\begin{algorithm}
        \begin{algorithmic}[1]
        \Procedure{registerAsSpecialInvoked}{$m$}
            \If{$mark(m.isSpecialInvoked)$}
                \State $t \gets m.declaringType$
                \State $t.specialInvokedMethods.add(m)$
                % \State
                \If{$t.instantiatedSubtypes.notEmpty()$}
                    \State $registerAsInvoked(m)$ 
                \EndIf
            \EndIf
        \EndProcedure
        % \State
        \Procedure{registerAsInstantiated}{$t$} 
            \State ...
            \For{$supert \in t.superTypes$}
            \State ...
                \For{$m \in supert.specialInvokedMethods$}
                    \State $registerAsInvoked(m)$ 
                \EndFor
            \EndFor
        \EndProcedure
        \end{algorithmic}
        \caption{RTA handling of invoke special.}
        \label{invokeSpecialAlg}
\end{algorithm}

%------------------------------------------------------------------------------
\subsection{Running Example}
%------------------------------------------------------------------------------

To demonstrate the idea of RTA with method summaries, consider
again the program in Figure~\ref{running-example}.
Summaries for all its methods can be found in Table
\ref{running-example-method-summaries}.
Note that the empty sets inside the summaries are omitted for brevity.
%
% All summaries are also put into a single table, whereas in an actual analysis
% run, 
%
We stress that the summaries presented in the table are in fact created lazily
when their corresponding methods are marked as invoked.

The method \texttt{Hello.main} is the entry point, therefore it is used to
initialize the worklist and consequently processed first.
Its bytecode is parsed and its summary is created.
As shown in the summary, it instantiates the types \texttt{Hello} and
\texttt{A}, has a virtual invoke of the method \texttt{Hello.foo}, and a direct
invoke of the method \texttt{Hello.log}. 

The summary for \texttt{Hello.main} is now applied to update the state of the
analysis.
First, the types \texttt{Hello} and \texttt{A} are marked as instantiated.
None of these types or their supertypes have any methods marked as virtually
invoked and no new call targets are discovered.
When processing the virtual method call \texttt{Hello.foo}, the set of all
instantiated subtypes of \texttt{Hello} is considered, which currently has only
one element, \texttt{Hello} itself.
The call is then resolved against the type \texttt{Hello}, which resolves to
\texttt{Hello.foo} as the call target.
Consequently, \texttt{Hello.foo} is marked as invoked and added into the
worklist.
The invoke of \texttt{Hello.log} is direct, so the corresponding method is also
marked as invoked and added to the worklist.
When the analysis of \texttt{Hello.main} finishes, there are two more methods to
process, \texttt{Hello.foo} and \texttt{Hello.log}. 

\begin{table}[tbp]
\caption{Method summaries for the running example.}
% \vspace{-5mm}
\begin{center}
\begin{tabular}{cccc}
\toprule
\textbf{Method}&\multicolumn{3}{c}{\textbf{Method summary}} \\
\cline{2-4} 
& \textbf{Instantiated types} & \multicolumn{2}{c}{\textbf{Method invokes} } \\
\cline{3-4} 
&  & \textbf{Direct}& \textbf{Virtual} \\
\midrule
\texttt{main} & \texttt{Hello}, \texttt{A}  & \texttt{log} & \texttt{foo} \\
\texttt{log} & \texttt{B}  &  &  \\
\texttt{foo} & &  & \texttt{I.bar} \\
\bottomrule
\end{tabular}
\label{running-example-method-summaries}
\end{center}
\end{table}

\begin{table}[tbp]
\caption{Results of the analyses on the running example.}
% \vspace{-5mm}
\begin{center}
\begin{tabular}{ccc}
\toprule
\textbf{Analysis}&\multicolumn{2}{c}{\textbf{Results}} \\
\cline{2-3} 
& \textbf{Instantiated types} & \textbf{Invoked methods}  \\
\midrule
\texttt{PTA} & \texttt{Hello}, \texttt{A}, \texttt{B}  & \texttt{log}, \texttt{foo}, \texttt{A.bar} \\
\texttt{RTA} & \texttt{Hello}, \texttt{A}, \texttt{B}  & \texttt{log}, \texttt{foo}, \texttt{\{A,{\color{red}B}\}.bar} \\
\bottomrule
\end{tabular}
\label{running-example-results}
\end{center}
\end{table}

\texttt{Hello.foo} virtually calls the method \texttt{I.bar}.
All instantiated subtypes of \texttt{I} are considered as receivers.
Currently, the only instantiated subtype of \texttt{I} is \texttt{A}.
Consequently, only \texttt{A.bar} is marked as invoked and added into the
worklist.

The analysis of \texttt{Hello.log} seems straightforward as it only marks the
type \texttt{B} as instantiated.
However, when traversing the supertypes of \texttt{B} in
\texttt{registerAsInstantiated}, the interface \texttt{I} is considered as well,
whose virtually invoked method \texttt{I.bar} is resolved against \texttt{B}.
This resolution identifies \texttt{B.bar} as a call target, which is then marked
as invoked and added into the worklist.
This is an example of a loss of precision compared to the points-to analysis,
which would correctly determine \texttt{A.bar} as the only call target. 
The results of running both analyses on the example are presented in Table~\ref{running-example-results}. 
The method \texttt{B.bar}, which is included among reachable
methods due to the imprecision of RTA, is highlighted in {\color{red}red}.

%------------------------------------------------------------------------------
\subsection{Heap Snapshotting and Embedded Constants}\label{heap_snapshotting}
%------------------------------------------------------------------------------

Application initialization at build time enables a significantly faster
application startup, but it poses a challenge for the analysis.
The initialization is executed already during analysis, when a given class is
marked as reachable.
The initialization code can create arbitrary objects and use them to initialize
static fields.
The object graphs reachable from these fields then have to be traversed because
they can contain types not seen in the analyzed methods.

The object graphs are traversed concurrently with the analysis by a component
called the \textit{heap scanner}.
The scanner works in tandem with the analysis and only processes the values of
fields that are marked as read.
Processing other fields is not necessary because if the analysis does
not discover any instruction that reads from a field, then its value can never
be read at runtime.
The scanner is notified by the analysis for every read field, and, if not
already done, it includes its content into
the image heap, and it also processes all objects transitively reachable from
the field's value by following its fields that are already marked as read.
If the heap scanner discovers a so-far unseen type, it notifies the analysis to treat it as
instantiated~\cite{native_image_paper}.

The values from static final fields of initialized classes can be constant
folded into the compiled methods during bytecode parsing.
We call such values \textit{embedded constants}.
Every time such a constant is discovered, it is given as a root to the heap
scanner.

To better explain the concept of constant folding of initialized static final
fields, take a look at the example in Figure~\ref{embedded-constants-example}.
Assume that the class \texttt{EmbeddedConstantsExample} is initialized at build
time, i.e., that the static initializer is executed during analysis.
The method \texttt{selectComponent} selects some component based on arbitrary
application logic.
The resulting object is used to initialize the field \texttt{c}.
The method \texttt{main} is the entry point.
When the analysis of \texttt{main} starts and its bytecode is parsed, the
compiler notices that the field access of \texttt{c} can be constant folded
because it was intialized and assigned a value that never changes (the field is
declared \texttt{final}).
Therefore the constant \texttt{c} is embedded into the compiler IR and then put
into the method summary.

%\vspace{-4mm}
\begin{figure}[htbp]
\begin{lstlisting}
public class EmbeddedConstantsExample {
    private static final Component c;
    static { c = selectComponent(); }
    private static Component selectComponent() {...}
    public static void main() { c.execute(); }
}
\end{lstlisting}
% \vspace{-5mm}
\caption{Embedded constants example.}
% \vspace{-4mm}
\label{embedded-constants-example}
\end{figure}

Assume that the method \texttt{selectComponent} is the only place where the
class \texttt{Component} is instantiated and this method is only reachable from
the class initializer of \texttt{EmbeddedConstantsExample}.
Without taking the embedded constant into consideration, the class
\texttt{Component} would not be considered as instantiated when processing the
virtual call of \texttt{Component.execute} and then its \texttt{execute} method
would not be considered as a call target, even though it is actually executed at
run time.
To handle this problem, the type of the embedded constant \texttt{c} and the
types of any other objects transitively reachable from the constant by following
fields marked as read are treated as instantiated.

%------------------------------------------------------------------------------
\subsection{Parallel Analysis}
%------------------------------------------------------------------------------

Algorithm~\ref{coreAlg} presented above is single-threaded.
To enable parallelism, we replace the explicit worklist with a parallel task
list (see Algorithm~\ref{paralelMarkInvokedAlg}).
Before the analysis is started, a thread pool is created, which executes all
scheduled tasks.
Every root method is passed immediately into \texttt{registerAsInvoked}
(line~2), which was updated in the following manner.
If the method \texttt{mark} returns true, the execution of \texttt{onInvoked} is
scheduled as a separate task (line~7) so that any available thread in the thread
pool can execute it.
The method \texttt{onInvoked} obtains the summary for each invoked method and
applies it to update the state of the analysis.

The methods \texttt{registerAsVirtualInvoked} and \texttt{registerAsInstantiated} 
of Algorithm~\ref{rtaMethodsAlg} do not need to be updated, both of them
call \texttt{registerAsInvoked}, which is already updated to be parallel.
The \texttt{mark} method of Algorithm~\ref{coreAlg} is already using an atomic
operation to ensure that only one thread processes a newly reachable element
even if multiple threads attempt to mark it concurrently.

Note that Algorithm~\ref{rtaMethodsAlg} is already carefully designed to be safe
with regards to parallel execution.
In \texttt{registerAsVirtualInvoked}, the method must be added to
\texttt{virtualInvokedMethods} (line~4) before iterating the instantiated
subtypes (lines~6--9).
Likewise, in \texttt{registerAsInstantiated}, the type must be added to all
\texttt{instantiatedSubtypes} sets (lines~15--17) before iterating the
\texttt{virtualInvokedMethods} (lines~19--24).
This guarantees that a concurrent execution of \texttt{instantiatedSubtypes} and
\texttt{virtualInvokedMethods} that affects the same virtual method does not
miss to mark any resolved methods. 
Indeed, regardless of whether the
virtual method is first marked as invoked or the type is first marked as
instantiated, the method is registered as invoked either by the loop on
line~6 in \texttt{registerAsVirtualInvoked} or the loop on line~20 in
\texttt{registerAsInstantiated}.
% \tv{I do not understand the below. Probably to be discussed: e.g., how is this
% related to the above? What are the reads and writes, which are mentioned below,
% in the above? How are the "happens-before edges between the initial writes and
% subsequent reads across the threads" formed?}
%
% \dk{There is a \textit{happens-before} relation between the initial writes and
% subsequent reads. Each thread produces two events that are ordered locally
% (within the thread): write and read, and happens-before edges will also be
% formed between the initial writes and subsequent reads across the threads
% depending on the actual order of execution.  There are only finitely many such
% orderings and in each of them, the analysis marks the appropriate method in the
% end.}
%
% \tv{What about the following simplification? Indeed, regardless of whether the
% virtual method is first marked as invoked or the type is first marked as
% instantiated, the method will be registered as invoked either by the loop on
% line~6 in \texttt{registerAsVirtualInvoked} or the loop on line~20 in
% \texttt{registerAsInstantiated}.}
%

% \vspace{-5mm}
\begin{algorithm}
        \begin{algorithmic}[1]
        
        \For{$m \in rootMethods$}
            \State $registerAsInvoked(m)$
        \EndFor
        % \State
        \Procedure{registerAsInvoked}{$m$}
            \If{$mark(m.isInvoked)$}
                \State $schedule(() \rightarrow onInvoked(m))$
            \EndIf
        \EndProcedure
        % \State
        \Procedure{onInvoked}{$m$}
            \State $irGraph \gets parseMethod(m)$
            \State $s \gets extractSummary(irGraph)$
            \State $applySummary(s)$ 
        \EndProcedure
        \end{algorithmic}
        \caption{Excerpts of the parallel analysis.}
        \label{paralelMarkInvokedAlg}
\end{algorithm}

%------------------------------------------------------------------------------
\subsection{Incremental Analysis} \label{IncrementalAnalysis}
%------------------------------------------------------------------------------

Method summaries are designed so that they can be easily serialized and reused.
Each method summary can be transformed into a purely textual
\texttt{SerializedSummary}.
Classes, methods, and fields are represented as follows:\begin{itemize}

  \item Each class is represented by a \texttt{ClassId}, which consists of the
  full name of the class.

  \item Each method is represented by a \texttt{MethodId} consisting of the
  \texttt{ClassId} of the declaring class, the method name, and the signature to
  differentiate overloaded methods.

  \item Each field is represented by a \texttt{FieldId} consisting of the
  \texttt{ClassId} of the declaring class and the field name.

\end{itemize}

The process of serializing summaries is straightforward because it only requires
to pick specific string identifiers based on the rules above.
On the other hand, the \textit{resolution}, which transforms the
\texttt{SerializedSummary} back into the \texttt{MethodSummary}, is more
complex. 

Resolving \texttt{ClassIds} back into classes is done by looking them up using a
specialized \texttt{Classloader}, which is a~special class responsible for
loading classes \cite{JVMSpec}.
Resolving methods and fields is a two-step process.
First, the declaring class is resolved.
If the class resolution is successful, the algorithm locates the
requested field/method by iterating over all declared methods/fields.
We aim to improve the lookup procedure in the future---the naive iteration is a
limitation of the current implementation only.

Unfortunately, not all summaries can be reused.
For a summary to be reusable, it has to match the following
criteria:\begin{itemize}

  \item Each identifier has to be \textit{stable}. We call an identifier stable
  if its resolution in different analysis runs always results in the same
  element. Unfortunately, lambda names, proxy names and in general names of all
  generated classes and methods are potentially unstable. 

  \item All embedded constants have to be \textit{trivial}. We call a constant
  trivial if it is a primitive data type or 
  %
  % a~commonly used % TV: whether a type is commonly used or nor should not play
  % any technical role here. Is so, one would also need a technical meaning of
  % being common, which I can't see how this would be done
  %
  an immutable type with a fixed internal structure, such as
  \texttt{java.lang.String}. If a~given class is immutable and has a fixed
  internal structure, the set of types in its object graph is identical for all
  instances. Therefore, it is enough to process only a single instance.  For
  commonly used types such as \texttt{java.lang.String}, it is guaranteed that
  at least one such instance is processed when traversing the image heap,
  and so these embedded constants can be ignored in summaries.

\end{itemize} Note, however, that both of these limitations are merely
implementation-specific.
They are not inherent to the proposed algorithm and could be lifted in
the future.
Implementing a proper handling for these two cases would be a significant
engineering effort with little added value research-wise---hence we
decided to keep these restrictions for now. 

To integrate the reuse of summaries into the previous algorithms, the process of
parsing the bytecode and extracting summaries is moved to a new procedure
\texttt{getSummmary} described in Algorithm~\ref{getSummaryAlg}.
The procedure first tries to load a serialized summary for the given method
(line~2).
At the moment, all serialized summaries preserved from previous compilations are
stored in a file, which is loaded into a map associating \texttt{MethodIds} to
corresponding serialized summaries.
However, the summaries could also be fetched from a remote source or included
with the libraries the compiled application is using, so that even the first
execution in a given context (user account, host, etc.) can benefit from
incrementality. 

Since the method could have changed in between the builds, it is important to
check validity of the summary (line~3).
That can be achieved by storing the hash of the bytecode instructions along with
the summary.
Smarter approaches could take into consideration timestamps on the jar files or
library version numbers, but since our goal was to estimate the benefit that can
be obtained by reusing summaries, we decided to use only hashing for the initial
prototype.

If the \texttt{SerializedSummary} is available and is still valid, it is
resolved back into a \texttt{MethodSummary} based on the rules described above
(line 4).
If the resolution is successful, the summary can be reused, otherwise it is
necessary to extract a new one by parsing the bytecode (lines 9--11). 

% \vspace{-4mm}
\begin{algorithm}%[t]
        \begin{algorithmic}[1]
        \Procedure{getSummary}{$method$}
            \State $serialized \gets loadSummary(method)$
            \If{$serialized \ne null\;and\;isValid(serialized)$}
                \State $summary \gets resolve(serialized)$
                \If{$summary \ne null$}
                    \State \Return $summary$
                \EndIf
            \EndIf
            \State $irGraph \gets parseMethod(method)$
            \State $summary \gets extractSummary(irGraph)$
            \State \Return $summary$
        \EndProcedure
        \end{algorithmic}
        \caption{Retrieving a method summary.}
        \label{getSummaryAlg}
\end{algorithm}
% \vspace{-4mm}

Reusing summaries from previous builds allows the analysis to skip the overhead
of parsing.
Unfortunately, parsing still has to occur for the compilation that follows, so
until the compilation pipeline is incremental as well, the benefits can be seen
only on the analysis time, not the whole build.

%==============================================================================
\section{Evaluation} \label{evaluation}
%==============================================================================

This section compares our implementation of RTA and PTA in the context of GraalVM Native Image.
We use Oracle GraalVM~23.0 based on JDK~20.

The experiments are executed on a dual-socket Intel Xeon E5-2630 v3 running at
2.40~GHz with 8~physical/16~logical cores per socket, 128~GiB main memory,
running Oracle Linux Server release~7.3.
The benchmark execution is pinned to one of the two CPUs, and TurboBoost was
disabled to avoid instability.
The number of threads is by default set to~16 with the exception of scalability
experiments where it is a part of the configuration.
Each benchmark is executed 10 times, and the average values are presented.
We do not include the deviation as it is significantly smaller than the differences between PTA and RTA in most cases.
We use the following applications for the evaluation:\begin{itemize}

\item \emph{Helloworld}: A simple Java application printing a text to the
standard output. Even such a simple application actually consists of more than
1,000 classes and 10,000 methods, e.g., for the necessary charset conversion
code and the runtime system.

\item \emph{DaCapo}: A benchmark suite that consists of client-side Java
benchmarks, trying to exercise the complex interactions between the
architecture, compiler, virtual machine and running application \cite{dacapo}.
We use a subset of the benchmark suite because some benchmarks are not
compatible with our AOT compilation.

\item \emph{Renaissance}: A benchmark suite that consists of  real-world,
concurrent, and object-oriented workloads that exercise various concurrency
primitives of the JVM \cite{renaissance}. We use a subset of the benchmark
suite because some benchmarks are not compatible with our AOT compilation.

\item \emph{\{Spring, Micronaut, Quarkus\} Helloworld}: Simple helloworld
applications in the corresponding frameworks.

\item \emph{Quarkus Registry}~\cite{QuarkusRegistry}: A large real-world
application using the Quarkus framework. It is used to host the Quarkus
extension registry.

\item \emph{Micronaut MuShop}~\cite{MuShop}: A large demo application using the
Micronaut framework. We use three services: Order, Payment, and User.

\item \emph{Spring
Petclinic}\footnote{https://github.com/spring-projects/spring-petclinic}: A
popular demo application of the Spring framework \cite{spring}.  

\item \emph{Micronaut Shopcart}: A demo application of the Micronaut framework
\cite{micronaut} performing similar tasks to the Petclinic but in a different
domain.

\item \emph{Quarkus Tika}\footnote{https://github.com/quarkiverse/quarkus-tika}:
An extension to the Quarkus Framework \cite{quarkus} that provides functionality
to parse documents using the Apache Tika
library\footnote{https://tika.apache.org/}.

\end{itemize}

The results are presented in Table~\ref{detailedTable}. The number of reachable
methods has been divided by 1,000 and similar conversions were performed to
present values in seconds and MB.
The values were rounded and then compared.
For Dacapo and Renaissance, the table presents a subset of the benchmarks only (a few small, a few mid-sized and a few of the biggest).
The data for all benchmarks can be found in the appendix\footnote{In case of acceptance, the paper including the appendix will be submitted to \url{https://arxiv.org/} so that the data are easily accessible.}.
We highlight Spring Petclinic in {\color{violet}violet} as we discuss its results often.

\begin{table*}[htb]
    \centering
    \caption{Detailed statistics of the evaluated benchmarks.}
    % \vspace{-5mm}
    \begin{tabular}{ll|rr|rr|rr|rr}
    \hline
        \toprule
        & & \multicolumn{2}{c}{Reachable Methods} & \multicolumn{2}{c}{Analysis Time [s]} & \multicolumn{2}{c}{Total time [s]} & \multicolumn{2}{c}{Binary size [MB]}\\
        Suite & Benchmark & PTA & RTA & PTA & RTA & PTA & RTA & PTA & RTA \\
        \midrule
\multirow{1}{*}{Console} & helloworld & 18 & \textbf{+17\%} & 14 & +21\% & 36 & \textbf{+17\%} & 13 & +23\% \\
\midrule
\multirow{4}{*}{Dacapo} & avrora & 24 & +25\% & 12 & -8\% & 51 & +6\% & 23 & +30\% \\
 % & batik & 52 & +12\% & 25 & -24\% & 80 & -2\% & 61 & +18\% \\
 & fop & 94 & +4\% & 46 & -30\% & 128 & -10\% & 105 & +11\% \\
 % & h2 & 40 & +8\% & 20 & -25\% & 69 & -4\% & 52 & +10\% \\
 & jython & 71 & +8\% & 55 & -35\% & 140 & -26\% & 134 & +9\% \\
 & luindex & 26 & +23\% & 13 & -8\% & 54 & +7\% & 32 & +25\% \\
 % & lusearch & 24 & +25\% & 12 & -8\% & 49 & +6\% & 29 & +28\% \\
 % & pmd & 62 & +5\% & 29 & -31\% & 90 & -7\% & 72 & +10\% \\
 % & sunflow & 52 & +12\% & 27 & -26\% & 88 & -2\% & 46 & +24\% \\
 % & xalan & 48 & +6\% & 24 & -29\% & 76 & -5\% & 48 & +15\% \\
\midrule
\multirow{9}{*}{Microservices} & micronaut-helloworld-wrk & 74 & +4\% & 34 & -32\% & 88 & -9\% & 45 & +18\% \\
 & mushop:order & 168 & +2\% & 102 & -59\% & 209 & -30\% & 104 & +13\% \\
 & mushop:payment & 82 & +4\% & 36 & -33\% & 91 & -10\% & 50 & +14\% \\
 & mushop:user & 115 & +3\% & 57 & -44\% & 135 & -18\% & 76 & +13\% \\
 & {\color{violet}petclinic-wrk} & {\color{violet}207} & {\color{violet}+4\%} & {\color{violet}159} & {\color{violet}\textbf{-64\%}} & {\color{violet}297} & {\color{violet}\textbf{-35\%}} & {\color{violet}144} & {\color{violet}\textbf{+15\%}} \\
 & quarkus-helloworld-wrk & 52 & +6\% & 18 & -22\% & 69 & -3\% & 50 & +4\% \\
 & quarkus:registry & 111 & +5\% & 49 & -39\% & 126 & -16\% & 69 & +19\% \\
 & spring-helloworld-wrk & 67 & +4\% & 30 & -33\% & 87 & -10\% & 47 & +13\% \\
 & tika-wrk & 82 & \textbf{+6\%} & 29 & -28\% & 117 & -6\% & 88 & \textbf{+6\%} \\
\midrule
% \multirow{23}{*}{Renaissance} & akka-uct & 35 & +17\% & 17 & 0\% & 49 & +2\% & 22 & +23\% \\
\multirow{8}{*}{Renaissance} & chi-square & 173 & \textbf{+8\%} & 129 & -60\% & 260 & -30\% & 100 & +17\% \\
 % & als & 317 & +6\% & 1558 & \textbf{-94\%} & X & X & X & X \\
 % & chi-square & 173 & \textbf{+8\%} & 129 & -60\% & 260 & -30\% & 100 & +17\% \\
 & dec-tree & 324 & +6\% & 2009 & \textbf{-95\%} & X & X & X & X \\
 % & dotty & 83 & +10\% & 37 & -16\% & 98 & -4\% & 51 & +16\% \\
 % & finagle-chirper & 89 & +8\% & 45 & -29\% & 105 & -12\% & 51 & +16\% \\
 % & finagle-http & 88 & +7\% & 48 & -33\% & 111 & -14\% & 50 & +18\% \\
 % & fj-kmeans & 26 & +23\% & 11 & 0\% & 35 & +6\% & 18 & +22\% \\
 & future-genetic & 27 & +22\% & 15 & 0\% & 44 & +5\% & 19 & +21\% \\
 & gauss-mix & 189 & +8\% & 146 & -61\% & 286 & -32\% & 107 & +17\% \\
 & log-regression & 334 & +7\% & 2215 & \textbf{-95\%} & X & X & X & X \\
 % & mnemonics & 26 & +23\% & 15 & 0\% & 43 & +2\% & 18 & +22\% \\
 % & movie-lens & 177 & +8\% & 109 & -59\% & 222 & -29\% & 106 & +15\% \\
 % & naive-bayes & 320 & +6\% & 1277 & -92\% & X & X & X & X \\
 & page-rank & 171 & +8\% & 129 & -60\% & 258 & -31\% & 119 & +13\% \\
 % & par-mnemonics & 27 & +19\% & 15 & 0\% & 43 & +5\% & 19 & +16\% \\
 % & philosophers & 28 & +18\% & 19 & +16\% & 48 & +10\% & 19 & +16\% \\
 & reactors & 30 & +13\% & 19 & +16\% & 47 & +11\% & 19 & +21\% \\
 % & rx-scrabble & 27 & +22\% & 14 & 0\% & 42 & +2\% & 27 & +15\% \\
 % & scala-doku & 27 & +22\% & 11 & 0\% & 35 & +6\% & 19 & +21\% \\
 % & scala-kmeans & 26 & +23\% & 14 & 0\% & 41 & +2\% & 18 & +22\% \\
 & scala-stm-bench7 & 30 & +20\% & 19 & +26\% & 49 & +14\% & 19 & +21\% \\
 % & scrabble & 27 & +19\% & 14 & 0\% & 41 & +5\% & 26 & +19\% \\

        \bottomrule
    \end{tabular}
    \label{detailedTable}
    % \vspace{-3mm}
\end{table*}

%------------------------------------------------------------------------------
\subsection{Reachable Elements}
%------------------------------------------------------------------------------

In order to get an insight into the actual size of our benchmarks, we measured
the number of reachable types, methods, and fields. Using metrics such as lines
of code or the number of classes could be misleading because only reachable
elements are analyzed and compiled.
Since these metrics are interconnected and follow the same pattern, we decided
to present the number of reachable methods as the main metric.
This number directly influences not only the scope of the analysis (how many
methods need to be processed) but also the workload of the compilation phase
afterwards. 
Details about types and fields can be found in the appendix.
%Reachable types and fields for each benchmark can be found in the appendix in Table~\ref{reachableProgramElements}.

We can immediately observe that the imprecision of RTA increases the number of
reachable methods for all benchmarks, as was expected.
However, an interesting trend can be observed.
Whereas there is a significant difference between reachable elements for
\texttt{HelloWorld} and the other smaller \texttt{Renaissance} and
\texttt{Dacapo} benchmarks, the difference gets usually significantly smaller
for the bigger applications.
Nevertheless, one cannot say that the difference is uniformly decreasing with
the increasing size of the applications.
Indeed, for example, the number of reachable methods for \texttt{Quarkus Tika}
is increased by \NUMBER{6}\,\%, while a much bigger \texttt{Renaissance
chi-square} is increased by \NUMBER{8}\,\%.
This suggests that not only the size of the compiled application but also its
structure influence the performance and precision.
% of the analysis.  

%------------------------------------------------------------------------------
\subsection{Analysis Time}
%------------------------------------------------------------------------------

The time that is reported by GraalVM Native Image as the analysis time includes
the time spent running application initialization code.
We treat this step as a constant factor that cannot be directly improved by
different analysis methods. In order to measure the influence on analysis more
precisely, we subtracted it from the overall analysis time.
It can be seen that RTA outperforms PTA on all benchmarks apart from the small
ones.
The most notable savings are for \texttt{Spring Petclinic}, for which the
analysis time is reduced by \NUMBER{64}\,\%.
The biggest \texttt{Renaissance} benchmarks
\texttt{log-regression}, and \texttt{dec-tree} also exhibit a significant
analysis time reduction.
Unfortunately, these benchmarks are not fully supported by GraalVM Native Image
and currently fail during compilation.  We have decided to include at least the
analysis time of these benchmarks because they are the biggest of our suite in
terms of reachable methods.  

%------------------------------------------------------------------------------
\subsection{Build Time}
%------------------------------------------------------------------------------

Since the reduced precision of RTA puts more workload on the compilation phase
that follows, we also measured the whole build time.
It can be seen that while the imprecision of RTA indeed negatively influences
small applications such as \texttt{HelloWorld} or smaller benchmarks from the
\texttt{Renaissance} and \texttt{Dacapo} bench suites, for bigger applications
the time saved in the analysis outweighs the extra compilation time.
The biggest savings were again obtained for \texttt{Spring Petclinic} where the
overall build was reduced by \NUMBER{35}\,\%. 

% \vspace{-3mm}
%------------------------------------------------------------------------------
\subsection{Binary Size}
%------------------------------------------------------------------------------

As another way to compare the precision of PTA and RTA, we measured the size of
the compiled image.
It can be seen that the size increases for all benchmarks and, in general, the size of smaller images 
increased more.
However, there does not seem to be a clear pattern. 
%
% However, the increase is not linear with respect to the size of the application
% in terms of reachable methods.
%
% For example, the binary size of \texttt{Quarkus Tika} was increased by
% \NUMBER{6}\,\% only, whereas the size of the much bigger \texttt{Spring
% Petclinic} was increased by \NUMBER{15}\,\%.
%
That can be attributed to the fact that the size of the image is influenced by
multiple factors (such as the metadata, embedded resources, etc.), not just the
results of the analysis.

% \vspace{-3mm}
%------------------------------------------------------------------------------
\subsection{Scalability with CPU Cores}
%------------------------------------------------------------------------------

To evaluate how PTA and RTA scale with the number of available CPU cores, we
executed each benchmark with 1, 4, 8, and 16 threads.
The results for several representative benchmarks are presented in Figure~\ref{fig:scalability-pta}, and Figure~\ref{fig:scalability-rta}, and the rest can be found in the appendix.

\begin{figure*}[h]
    \begin{subfigure}{.49\textwidth}
        \includegraphics[scale=0.17]{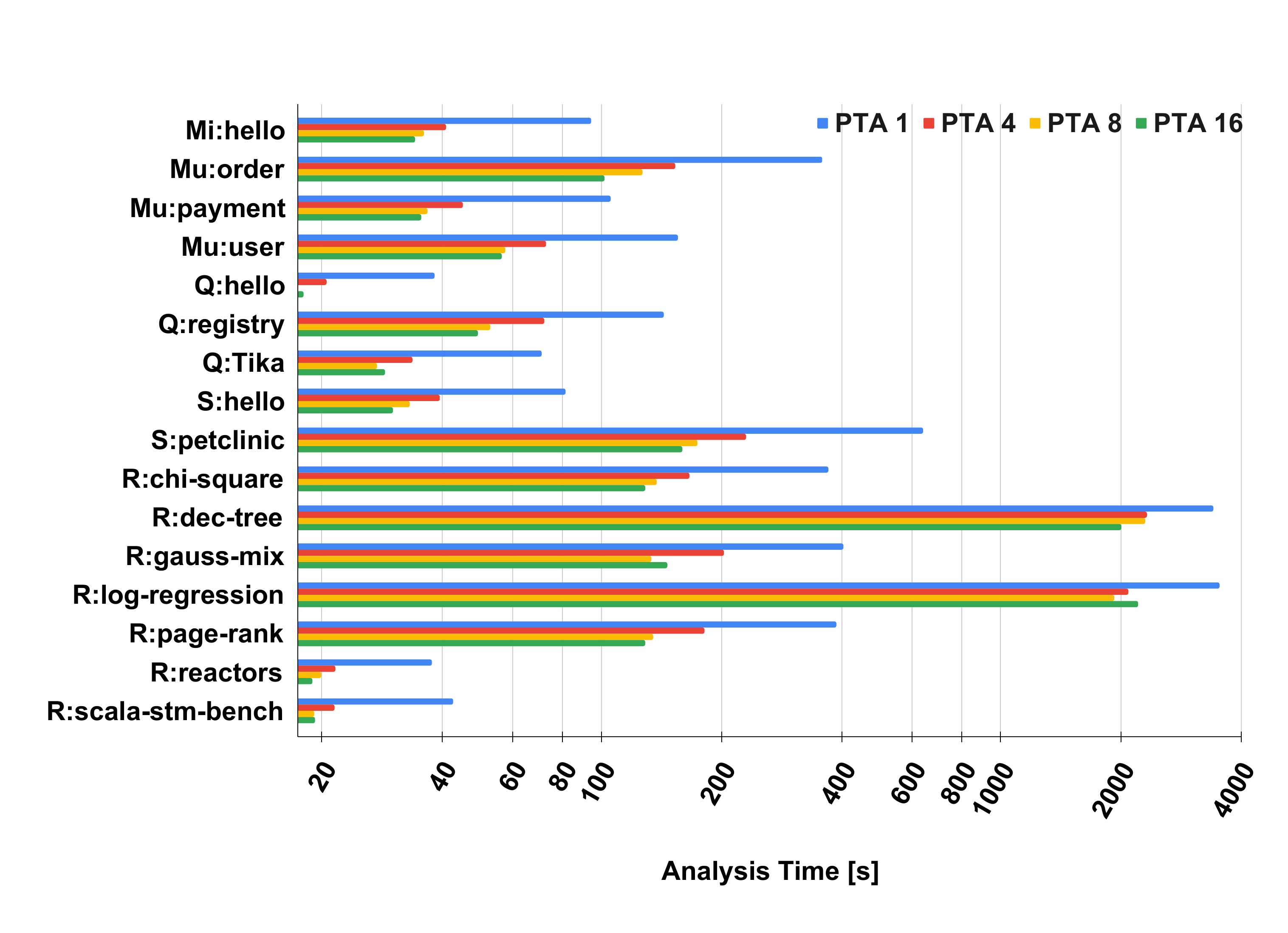}
        \centering
        % \vspace{-5mm}
        \caption{Scalability PTA results.}
        \label{fig:scalability-pta}
    \end{subfigure}\quad
    \begin{subfigure}{.49\textwidth}
        \includegraphics[scale=0.17]{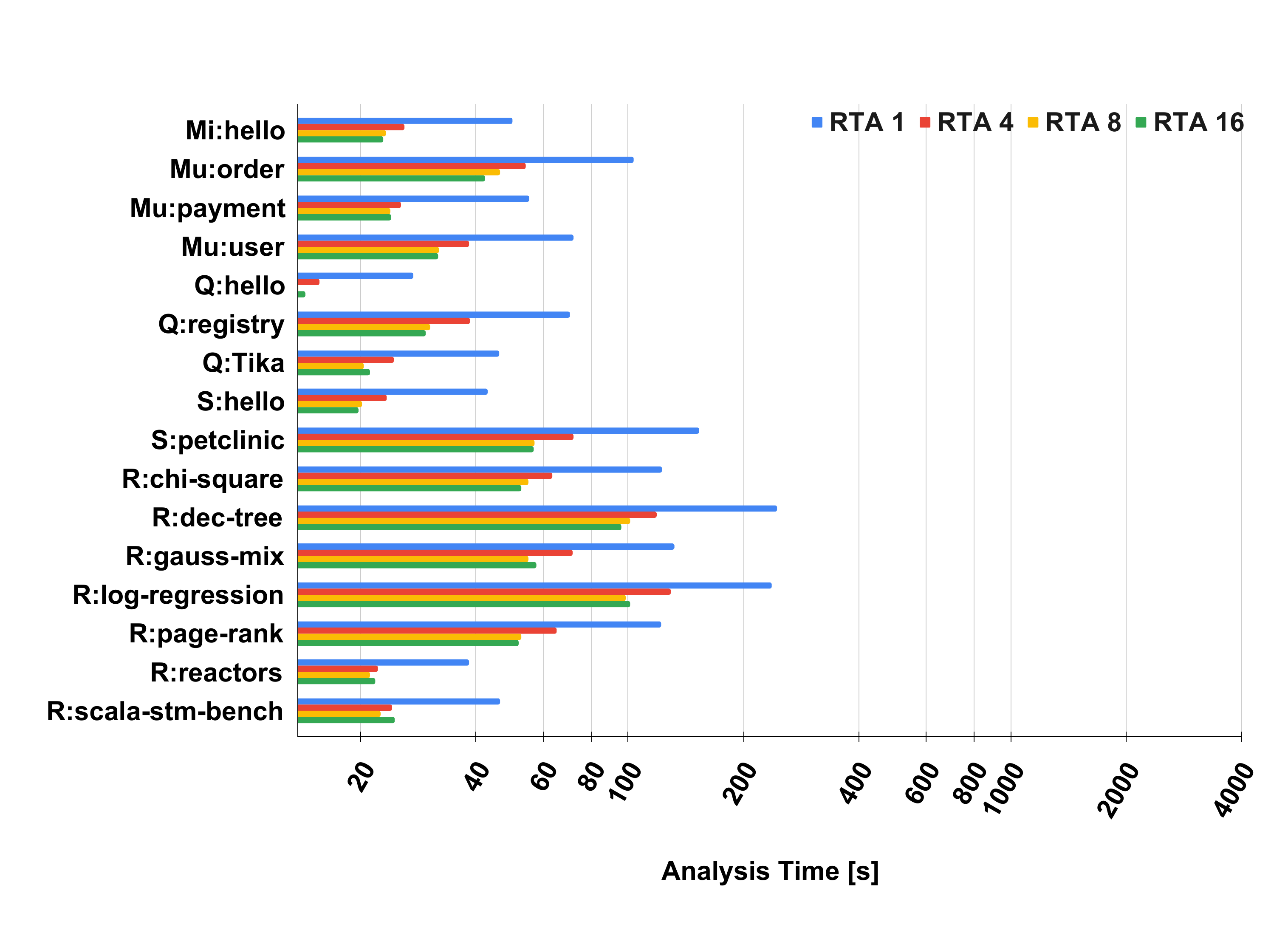}
        \centering
        % \vspace{-5mm}
        \caption{Scalability RTA results.}
        \label{fig:scalability-rta}
    \end{subfigure}
    % \vspace{-5mm}
    \caption{Scalability results (log scale).}
    % \vspace{-5mm}
    \label{fig:scalability}
\end{figure*}

% \begin{figure}[h]
% \includegraphics[width=0.45\textwidth]{chart-pta.png}
% \centering
% \caption{Scalability PTA results.}
% \label{fig:scalability-pta}
% \end{figure}

% \begin{figure}[h]
% \includegraphics[width=0.45\textwidth]{chart-rta.png}
% \centering
% \caption{Scalability RTA results.}
% \label{fig:scalability-rta}
% \end{figure}

%
By looking at the figures, it can be seen that RTA outperformed PTA in most
experiments and performed especially well in scenarios with a reduced number of
threads.
For example, the analysis time of Spring Petclinic using only a single thread
was reduced by \NUMBER{76}\;\%. 

Conversely, as the number of threads increases, the difference is
reduced
%
% , which  is a trend that can be observed
% 
in most benchmarks.
It suggests that the current implementation of RTA might contain some
scalability bottlenecks.
While the implementation of PTA is production-ready and has been optimized for
many years, our implementation of RTA is still a research prototype.
Therefore, the existence of such scalability bottlenecks is not surprising and
suggests that even better results could be achieved if more time is invested
into profiling and optimization of the analysis. 

%------------------------------------------------------------------------------
\subsection{Runtime Performance}
%------------------------------------------------------------------------------

Since our implementation of RTA is meant for the development mode and not
for production deployments, we focused mainly on the build-time characteristics. 
In spite of that, we have also collected runtime data for \texttt{Renaissance}
and \texttt{Dacapo} to provide a more complete picture. For space reasons, we
provide only aggregated statistics. 
We observed that the time to execute a standard workload for the benchmarks was
increased on average by \NUMBER{10}\;\% across all benchmarks, \NUMBER{11}\;\%
for \texttt{Reinaissance}, and \NUMBER{8}\;\% for \texttt{Dacapo}. 
The biggest increase was \NUMBER{26}\;\% for the \texttt{Renaissance
future-genetic} benchmark.

Even though such an increase is non-negligible, as already said, RTA is meant to
be used for development and testing where such a performance decrease is
justified by the reduced build time. 
If runtime performance is an important criterion, PTA should be considered
instead.

% \vspace{-3mm}
%------------------------------------------------------------------------------
\subsection{Incrementality}
%------------------------------------------------------------------------------

We have implemented the approach described in Section~\ref{IncrementalAnalysis}
and noted that more than \NUMBER{60}\,\% of summaries can be reused for
\texttt{Spring Petclinic}, one of the biggest benchmarks, because they satisfy
the necessary requirements.
Unfortunately, no real benefits were visible when reusing them.
It turned out that even though \NUMBER{60}\,\% of the methods did not have to be
parsed, parsing these methods only constituted about \NUMBER{33}\,\% of the
overall parse time.
The methods that would benefit from incrementality the most in our benchmarks
are unfortunately the same methods that contain non-trivial embedded constants.
As we discussed in Section~\ref{IncrementalAnalysis}, both of these limitations
are only implementation specific.
They are not inherent to the proposed algorithm and as it turned out that they are 
blocking the benefits.
% Therefore, we decided to lift the requirements for reusing method summaries.
% The results then present only an estimate about how much can be saved. We plan
% to resolve the issue with embedded constants properly in the follow up work. 

% \subsection{Implementation} \todo{add some klocs of code and developer time
% measurements} We also considered the engineering effort necessary to implement
% PTA and RTA in Native Image. Since significantly less work is necessary to
% implement RTA, we suggest using it as the first iteration in the
% implementation of AOT compiler and only returning to PTA if the imprecision
% becomes an issue. 

%==============================================================================
\section{Related Work} \label{state_of_the_art}
%==============================================================================

In \cite{practical_method_resolution_java,fast_analysis_cpp}, the authors
described multiple different approaches (including RTA) on how to construct the application call
graph, which is a necessary step for computing reachable program elements.
Our approach is an extension of RTA,
which is designed to be parallel, incremental, and also provides support for
heap snapshotting, a feature necessary for enabling class initialization at
build time.
In \cite{practical_method_resolution_java}, the authors also experimented with
Variable-Type Analysis, which seems to be similar to the points-to analysis used
in GraalVM Native Image.
Both works provided only a simple textual description of rapid type
analysis without any pseudocode.
On the contrary, we provide pseudocode and detailed description for all key
components. 

Tip and Palsberg gave an overview of various propagation-based call-graph
construction algorithms, again including RTA, in \cite{scalablePropagationBasedCallGraphAlgs}.
Using the terminology from their article, the points-to analysis in Native Image
could be classified as 0-CFA.
The authors also introduced four new algorithms CTA, FTA, MTA, and XTA that lie
between RTA and 0-CFA in the design space.
Based on the experimental evaluation, they concluded that their new algorithms,
while in theory more powerful than RTA, have only a minor effect with regards to
the number of reachable elements (up to less 3\,\% reachable methods), while
being up to 8.3 times slower than RTA.
On the other hand, the amount of call graph edges and uniquely resolved
polymorphic call sites can be reduced by up to 29\,\% and 26.3\,\%,
respectively.
Since the goal of our research was to reduce the analysis time, and performance
is not a priority in development builds, RTA seems to fit our use case best.
% be the best fit for our use case.    

% we build on similar principles as ... and we extend them with
%
In \cite{virgil}, B. Titzer proposed the \textit{Reachable Method Analysis},
which is similar to our core algorithm presented in Algorithm~\ref{coreAlg}. 
Our contributions on top of his analysis are using method summaries, incremental
approach, and experimental evaluation of PTA and RTA in the context of Native
Image.

Even though the analysis implemented in GraalVM Native Image is
context-insensitive and contains several optimizations which aim to increase
scalability by sacrificing precision \cite{native_image_paper}, the analysis can
still take minutes for bigger applications.
Our version of RTA can reduce the analysis time by up
to \NUMBER{64}\,\%.

Grech et al. used heap snapshots to improve the performance and precision of
whole program pointer analysis \cite{shootingFromTheHeap}.
However, their analysis is intentionally incomplete; It might miss some
reachable program elements.
Unfortunately, this is unacceptable for Native Image because if a method that
was not marked reachable by static analysis is executed at runtime, it is a
fatal error.

There are several tools that compile JVM-based languages into native binaries.
Kotlin Native \cite{kotlinNative} and Scala Native \cite{scalaNative} are two
examples of such.
However, both of them support only a specific language.
We support any language that can be compiled into JVM
bytecode.
Also, the analysis they use to determine reachable elements is not clearly
specified.

The OVM Real-time Java VM \cite{ovmPaper} AOT compiles Java applications into
executable images.
The OVM compiler uses an analysis method called Reaching Types Analysis to
detect what parts of code are reachable, but the authors do not specify any
details about the analysis in the paper. 

% \vspace{-5mm}
%==============================================================================
\section{Conclusions} \label{conclusion}
%==============================================================================

In this paper, we have introduced a new variant of rapid type analysis (RTA), which is
parallel, incremental, and supports heap snapshotting.
The incrementality is enabled by the use of method summaries, which can be
serialized and reused between multiple builds.
We have described the analysis by providing pseudocode for all key components. 

The analysis was implemented and evaluated in the context of GraalVM Native
Image.
RTA was then compared against the poins-to analysis currently
used in GraalVM Native Image.
We used the Java benchmark suites Renaissance and Dacapo along with example
applications for the mainstream Java microservice frameworks Spring, Micronaut,
and Quarkus.
The experimental evaluation showed, e.g., that RTA can reduce
the analysis time of the Spring Petclinic demo application by~\NUMBER{64}\,\% at
the cost of increasing the image size~by~\NUMBER{15}\,\%. 

We also experimented with the scalability of both our RTA and points-to analysis
wrt. the number of processor cores showing that, for a reduced number of threads
such as 1 or 4, the savings in the analysis time can be even greater, making RTA
a good choice for constrained environments such as GitHub Actions or similar CI
pipelines.

In the future, we plan to lift the restrictions currently imposed on which
method summaries can be reused.
On top of that, we plan to extend the incremental analysis by a concept of
\emph{summary aggregation} whose goal is to merge summaries of directly
connected methods.
Fewer but larger summaries should be beneficial when method summaries are
serialized and reused, boosting the effect of incrementality. 

\begin{acks}

This work has been supported by the Czech Science Foundation project 23-06506S
and the FIT BUT internal project FIT-S-23-8151.
We thank all members of the GraalVM team at Oracle Labs and the Institute for System Software at the Johannes Kepler University Linz for their support and contributions.

Oracle and Java are registered trademarks of Oracle and/or its affiliates.
Other names may be trademarks of their respective owners.

\end{acks}

\bibliographystyle{ACM-Reference-Format}
%\bibliography{main}

\appendix

%==============================================================================
\section{Detailed Results}
%==============================================================================
Results for all our benchmarks are presented in Table~\ref{detailedTableAll}.

\begin{table*}[h]
    \centering
    \caption{Detailed statistics of the evaluated benchmarks.}
    \begin{tabular}{ll|rr|rr|rr|rr}
    \hline
        \toprule
        & & \multicolumn{2}{c}{Reachable Methods} & \multicolumn{2}{c}{Analysis Time [s]} & \multicolumn{2}{c}{Total time [s]} & \multicolumn{2}{c}{Binary size [MB]}\\
        Suite & Benchmark & PTA & RTA & PTA & RTA & PTA & RTA & PTA & RTA \\
        \midrule
\multirow{1}{*}{Console} & helloworld & 18 & \textbf{+17\%} & 14 & +21\% & 36 & \textbf{+17\%} & 13 & +23\% \\
\midrule
\multirow{10}{*}{Dacapo} & avrora & 24 & +25\% & 12 & -8\% & 51 & +6\% & 23 & +30\% \\
 & batik & 52 & +12\% & 25 & -24\% & 80 & -2\% & 61 & +18\% \\
 & fop & 94 & +4\% & 46 & -30\% & 128 & -10\% & 105 & +11\% \\
 & h2 & 40 & +8\% & 20 & -25\% & 69 & -4\% & 52 & +10\% \\
 & jython & 71 & +8\% & 55 & -35\% & 140 & -26\% & 134 & +9\% \\
 & luindex & 26 & +23\% & 13 & -8\% & 54 & +7\% & 32 & +25\% \\
 & lusearch & 24 & +25\% & 12 & -8\% & 49 & +6\% & 29 & +28\% \\
 & pmd & 62 & +5\% & 29 & -31\% & 90 & -7\% & 72 & +10\% \\
 & sunflow & 52 & +12\% & 27 & -26\% & 88 & -2\% & 46 & +24\% \\
 & xalan & 48 & +6\% & 24 & -29\% & 76 & -5\% & 48 & +15\% \\
\midrule
\multirow{9}{*}{Microservices} & micronaut-helloworld-wrk & 74 & +4\% & 34 & -32\% & 88 & -9\% & 45 & +18\% \\
 & mushop:order & 168 & +2\% & 102 & -59\% & 209 & -30\% & 104 & +13\% \\
 & mushop:payment & 82 & +4\% & 36 & -33\% & 91 & -10\% & 50 & +14\% \\
 & mushop:user & 115 & +3\% & 57 & -44\% & 135 & -18\% & 76 & +13\% \\
 & {\color{violet}petclinic-wrk} & {\color{violet}207} & {\color{violet}+4\%} & {\color{violet}159} & {\color{violet}\textbf{-64\%}} & {\color{violet}297} & {\color{violet}\textbf{-35\%}} & {\color{violet}144} & {\color{violet}\textbf{+15\%}} \\
 & quarkus-helloworld-wrk & 52 & +6\% & 18 & -22\% & 69 & -3\% & 50 & +4\% \\
 & quarkus:registry & 111 & +5\% & 49 & -39\% & 126 & -16\% & 69 & +19\% \\
 & spring-helloworld-wrk & 67 & +4\% & 30 & -33\% & 87 & -10\% & 47 & +13\% \\
 & tika-wrk & 82 & \textbf{+6\%} & 29 & -28\% & 117 & -6\% & 88 & \textbf{+6\%} \\
\midrule
\multirow{23}{*}{Renaissance} & akka-uct & 35 & +17\% & 17 & 0\% & 49 & +2\% & 22 & +23\% \\
 & als & 317 & +6\% & 1558 & \textbf{-94\%} & X & X & X & X \\
 & chi-square & 173 & \textbf{+8\%} & 129 & -60\% & 260 & -30\% & 100 & +17\% \\
 & dec-tree & 324 & +6\% & 2009 & \textbf{-95\%} & X & X & X & X \\
 & dotty & 83 & +10\% & 37 & -16\% & 98 & -4\% & 51 & +16\% \\
 & finagle-chirper & 89 & +8\% & 45 & -29\% & 105 & -12\% & 51 & +16\% \\
 & finagle-http & 88 & +7\% & 48 & -33\% & 111 & -14\% & 50 & +18\% \\
 & fj-kmeans & 26 & +23\% & 11 & 0\% & 35 & +6\% & 18 & +22\% \\
 & future-genetic & 27 & +22\% & 15 & 0\% & 44 & +5\% & 19 & +21\% \\
 & gauss-mix & 189 & +8\% & 146 & -61\% & 286 & -32\% & 107 & +17\% \\
 & log-regression & 334 & +7\% & 2215 & \textbf{-95\%} & X & X & X & X \\
 & mnemonics & 26 & +23\% & 15 & 0\% & 43 & +2\% & 18 & +22\% \\
 & movie-lens & 177 & +8\% & 109 & -59\% & 222 & -29\% & 106 & +15\% \\
 & naive-bayes & 320 & +6\% & 1277 & -92\% & X & X & X & X \\
 & page-rank & 171 & +8\% & 129 & -60\% & 258 & -31\% & 119 & +13\% \\
 & par-mnemonics & 27 & +19\% & 15 & 0\% & 43 & +5\% & 19 & +16\% \\
 & philosophers & 28 & +18\% & 19 & +16\% & 48 & +10\% & 19 & +16\% \\
 & reactors & 30 & +13\% & 19 & +16\% & 47 & +11\% & 19 & +21\% \\
 & rx-scrabble & 27 & +22\% & 14 & 0\% & 42 & +2\% & 27 & +15\% \\
 & scala-doku & 27 & +22\% & 11 & 0\% & 35 & +6\% & 19 & +21\% \\
 & scala-kmeans & 26 & +23\% & 14 & 0\% & 41 & +2\% & 18 & +22\% \\
 & scala-stm-bench7 & 30 & +20\% & 19 & +26\% & 49 & +14\% & 19 & +21\% \\
 & scrabble & 27 & +19\% & 14 & 0\% & 41 & +5\% & 26 & +19\% \\

        \bottomrule
    \end{tabular}
    \label{detailedTableAll}
\end{table*}

%==============================================================================
\section{Detailed Reachable Program Elements}
%==============================================================================

Reachable program elements (types, methods, and fields) computed by points-to
analysis and our RTA for the benchmarks presented in Section~\ref{evaluation}
are presented in detail in Table~\ref{reachableProgramElements}.
As with the previous tables, the number of reachable program elements has been
divided by 1,000 and rounded before comparison.
By inspecting the table, it can be seen that these three metrics are correlated.
Benchmarks with more reachable methods have also more reachable types and fields
and vice versa.

\begin{table*}[h]
    \centering
    \caption{Reachable program elements (in thousands).}
    \begin{tabular}{ll|rrr|rrr|rrr}
    \hline
        \toprule
        & & \multicolumn{3}{c}{Reachable Types} & \multicolumn{3}{c}{Reachable Methods} & \multicolumn{3}{c}{Reachable Fields} \\
        Suite & Benchmark & PTA & RTA & Diff [\%] & PTA & RTA & Diff [\%] & PTA & RTA & Diff [\%] \\
        \midrule
\multirow{1}{*}{Console} & helloworld & 4 & 4 & 0\% & 18 & 21 & +17\% & 4 & 5 & +25\% \\
\midrule
\multirow{10}{*}{Dacapo} & avrora & 5 & 6 & +20\% & 24 & 30 & +25\% & 7 & 8 & +14\% \\
 & batik & 9 & 10 & +11\% & 52 & 58 & +12\% & 18 & 20 & +11\% \\
 & fop & 15 & 16 & +7\% & 94 & 98 & +4\% & 32 & 32 & 0\% \\
 & h2 & 7 & 7 & 0\% & 40 & 43 & +8\% & 11 & 12 & +9\% \\
 & jython & 10 & 11 & +10\% & 71 & 77 & +8\% & 18 & 19 & +6\% \\
 & luindex & 5 & 6 & +20\% & 26 & 32 & +23\% & 8 & 9 & +12\% \\
 & lusearch & 4 & 6 & +50\% & 24 & 30 & +25\% & 7 & 8 & +14\% \\
 & pmd & 10 & 11 & +10\% & 62 & 65 & +5\% & 19 & 20 & +5\% \\
 & sunflow & 9 & 10 & +11\% & 52 & 58 & +12\% & 18 & 20 & +11\% \\
 & xalan & 8 & 8 & 0\% & 48 & 51 & +6\% & 15 & 16 & +7\% \\
\midrule
\multirow{9}{*}{Microservices} & micronaut-helloworld-wrk & 13 & 14 & +8\% & 74 & 77 & +4\% & 19 & 19 & 0\% \\
 & mushop:order & 29 & 30 & +3\% & 168 & 172 & +2\% & 48 & 49 & +2\% \\
 & mushop:payment & 15 & 16 & +7\% & 82 & 85 & +4\% & 21 & 21 & 0\% \\
 & mushop:user & 20 & 21 & +5\% & 115 & 118 & +3\% & 31 & 31 & 0\% \\
 & {\color{violet}petclinic-wrk} & {\color{violet}39} & {\color{violet}40} & {\color{violet}+3\%} & {\color{violet}207} & {\color{violet}216} & {\color{violet}+4\%} & {\color{violet}65} & {\color{violet}66} & {\color{violet}+2\%} \\
 & quarkus-helloworld-wrk & 11 & 11 & 0\% & 52 & 55 & +6\% & 15 & 16 & +7\% \\
 & quarkus:registry & 20 & 21 & +5\% & 111 & 117 & +5\% & 28 & 29 & +4\% \\
 & spring-helloworld-wrk & 12 & 13 & +8\% & 67 & 70 & +4\% & 19 & 19 & 0\% \\
 & tika-wrk & 16 & 16 & 0\% & 82 & 87 & +6\% & 27 & 28 & +4\% \\
\midrule
\multirow{23}{*}{Renaissance} & akka-uct & 7 & 7 & 0\% & 35 & 41 & +17\% & 8 & 9 & +12\% \\
 & als & 57 & 58 & +2\% & 317 & 337 & +6\% & 85 & 86 & +1\% \\
 & chi-square & 30 & 31 & +3\% & 173 & 186 & +8\% & 51 & 51 & 0\% \\
 & dec-tree & 58 & 59 & +2\% & 324 & 344 & +6\% & 87 & 88 & +1\% \\
 & dotty & 15 & 16 & +7\% & 83 & 91 & +10\% & 25 & 26 & +4\% \\
 & finagle-chirper & 16 & 17 & +6\% & 89 & 96 & +8\% & 22 & 23 & +5\% \\
 & finagle-http & 16 & 17 & +6\% & 88 & 94 & +7\% & 22 & 22 & 0\% \\
 & fj-kmeans & 5 & 6 & +20\% & 26 & 32 & +23\% & 6 & 7 & +17\% \\
 & future-genetic & 5 & 6 & +20\% & 27 & 33 & +22\% & 6 & 7 & +17\% \\
 & gauss-mix & 32 & 34 & +6\% & 189 & 204 & +8\% & 54 & 54 & 0\% \\
 & log-regression & 59 & 60 & +2\% & 334 & 357 & +7\% & 88 & 89 & +1\% \\
 & mnemonics & 5 & 6 & +20\% & 26 & 32 & +23\% & 6 & 7 & +17\% \\
 & movie-lens & 31 & 32 & +3\% & 177 & 191 & +8\% & 51 & 52 & +2\% \\
 & naive-bayes & 57 & 59 & +4\% & 320 & 339 & +6\% & 86 & 87 & +1\% \\
 & page-rank & 30 & 31 & +3\% & 171 & 184 & +8\% & 49 & 50 & +2\% \\
 & par-mnemonics & 5 & 6 & +20\% & 27 & 32 & +19\% & 6 & 7 & +17\% \\
 & philosophers & 5 & 6 & +20\% & 28 & 33 & +18\% & 6 & 7 & +17\% \\
 & reactors & 6 & 6 & 0\% & 30 & 34 & +13\% & 7 & 7 & 0\% \\
 & rx-scrabble & 5 & 6 & +20\% & 27 & 33 & +22\% & 6 & 7 & +17\% \\
 & scala-doku & 5 & 6 & +20\% & 27 & 33 & +22\% & 6 & 7 & +17\% \\
 & scala-kmeans & 5 & 6 & +20\% & 26 & 32 & +23\% & 6 & 7 & +17\% \\
 & scala-stm-bench7 & 6 & 6 & 0\% & 30 & 36 & +20\% & 7 & 7 & 0\% \\
 & scrabble & 5 & 6 & +20\% & 27 & 32 & +19\% & 6 & 7 & +17\% \\

        \bottomrule
    \end{tabular}
    \label{reachableProgramElements}
\end{table*}

\section{Detailed Scalability Results}
\begin{figure*}[h]
\includegraphics[width=0.93\textwidth]{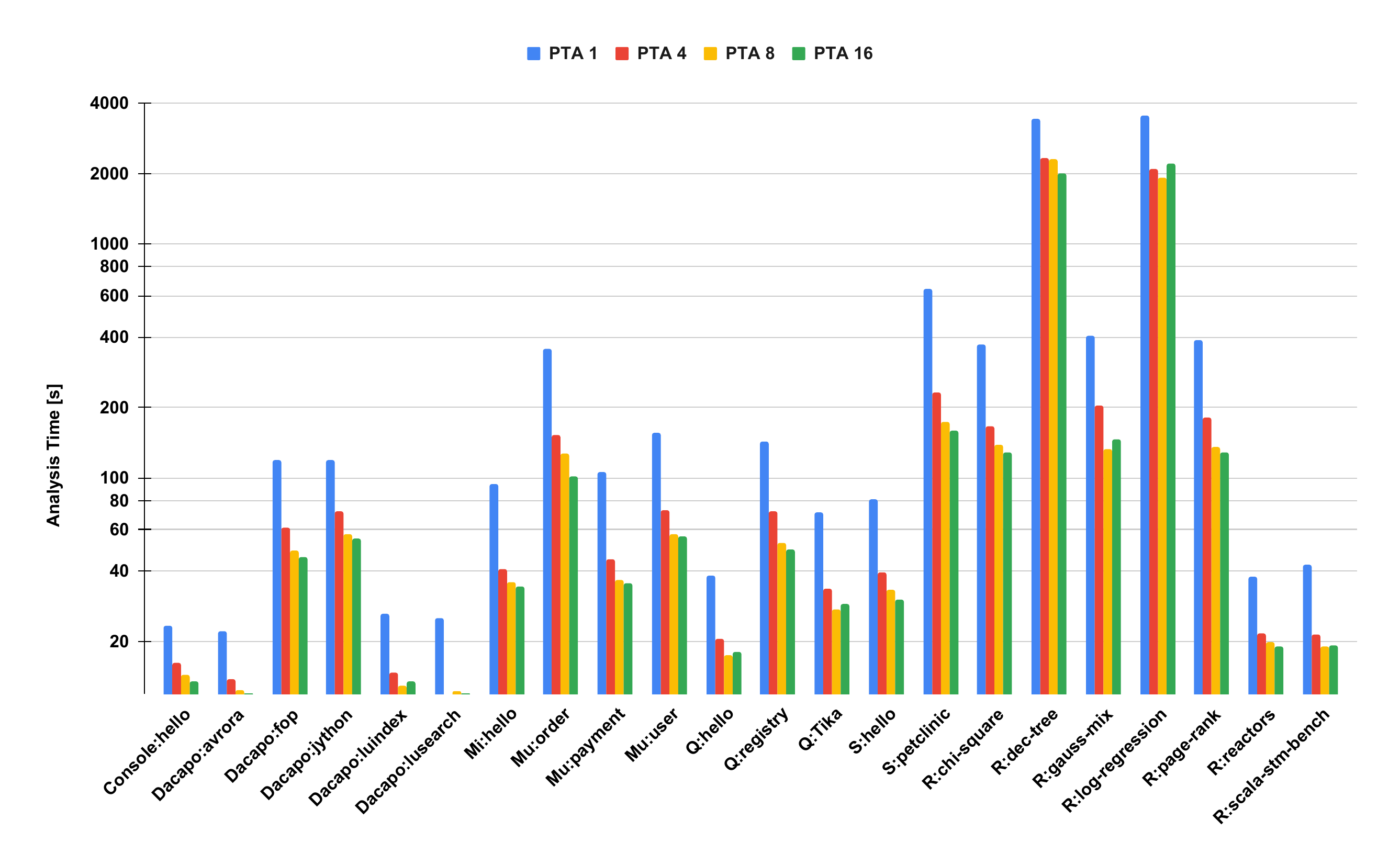}
\centering
\caption{Scalability PTA results (log scale).}
\label{fig:scalability-pta-all}
\end{figure*}

\begin{figure*}[h]
\includegraphics[width=0.93\textwidth]{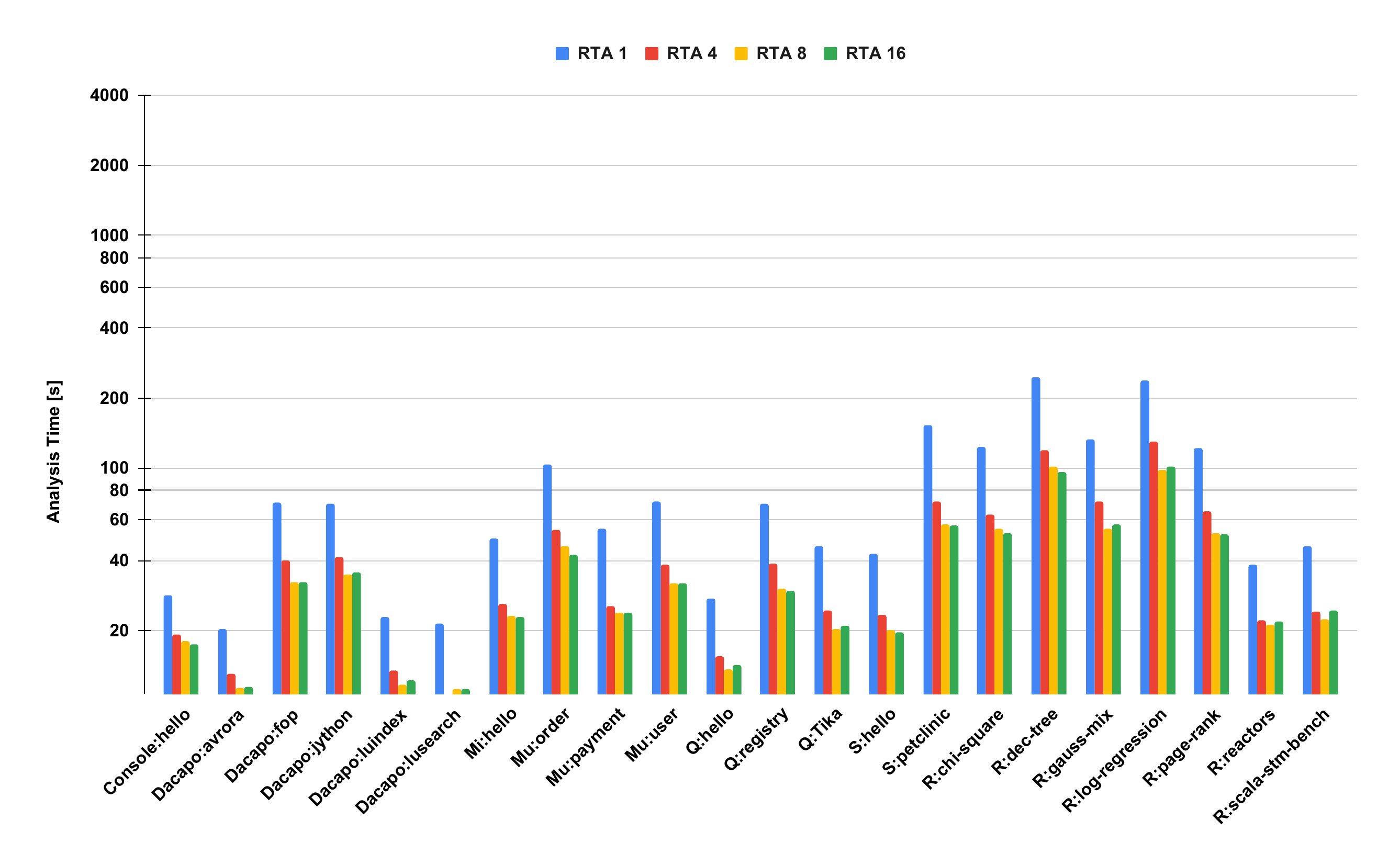}
\centering
\caption{Scalability RTA results (log scale).}
\label{fig:scalability-rta-all}
\end{figure*}

Detailed scalability results are presented in Figure~\ref{fig:scalability-pta-all} and Figure~\ref{fig:scalability-rta-all}.

\end{document}